\documentclass[12pt,a4paper]{article}
\usepackage{graphicx}
\usepackage{color}
\usepackage{citesort}
\usepackage{latexsym}
\usepackage{amsmath}
\usepackage{amssymb}

\newcommand{\vect}[1]{\mbox{\boldmath $#1$}}
\newcommand{\ie}{\textit{i.e.}}
\newcommand{\eg}{\textit{e.g.}}
\newcommand{\Prob}{{\rm Prob}}
\newcommand{\FIG}{figure~}

\newcommand{\TAB}{table~}

\newcommand{\SEC}{section~}

\begin{document}

\title{Importance of individual events in temporal networks}
	
\author{Taro Takaguchi$^1$, Nobuo Sato$^2$, Kazuo Yano$^2$, and Naoki Masuda$^{1,3,*}$
\\
\\
\\
${}^{1}$ 
Department of Mathematical Informatics,\\
The University of Tokyo,\\
7-3-1 Hongo, Bunkyo, Tokyo 113-8656, Japan
\\
\\
${}^{2}$
Central Research Laboratory, Hitachi, Ltd.,\\
1-280 Higashi-Koigakubo, Kokubunji-shi, Tokyo 185-8601, Japan
\\
\\
${}^{3}$
PRESTO, Japan Science and Technology Agency,\\
4-1-8 Honcho, Kawaguchi, Saitama 332-0012, Japan\\
\\
* Corresponding author (masuda@mist.i.u-tokyo.ac.jp)
}

\setlength{\baselineskip}{0.77cm}
\maketitle

\newpage

\begin{abstract}
\setlength{\baselineskip}{0.77cm}
Records of time-stamped social interactions between pairs of individuals
(\eg,  face-to-face conversations, e-mail exchanges, and phone calls) constitute a so-called temporal network.
A remarkable difference between temporal networks and conventional static networks is
that time-stamped events rather than links are the unit elements generating the collective behavior of nodes.
We propose an importance measure for single interaction events.
By generalizing the concept of the advance of event proposed by [Kossinets~G, Kleinberg~J and Watts~D~J 2008 {\em Proceeding of the 14th ACM SIGKDD International conference on knowledge discovery and data mining} p~435],
we propose that an event is central when it carries new information about others to the two nodes involved in the event. 
We find that the proposed measure properly quantifies the importance of events in connecting nodes along time-ordered paths.
Because of strong heterogeneity in the importance of events present in real data,
a small fraction of highly important events is necessary and sufficient to sustain the connectivity of temporal networks.
Nevertheless, in contrast to the behavior of scale-free networks against link removal,
this property mainly results from bursty activity patterns and not heterogeneous degree distributions.
\end{abstract}

\newpage

\section{Introduction}
Development of sensor technologies and the prevalence of electronic communication services
provide us with massive amount on data of human communication behavior,
including face-to-face conversations~\cite{Takaguchi2011,Cattuto2010,Isella2011}, e-mail exchanges~\cite{Eckmann2004,Barabasi2005,Malmgren2008}, phone calls~\cite{Onnela2007,Gonzalez2008,Candia2008},
and message exchanges and other types of interactions in various online forums~\cite{Holme2004,Rocha2010,Szell2010}.
Such data are collectively referred to as temporal networks, where time-stamped events, rather than static links,
are assumed between pairs of nodes (\ie, individuals)~\cite{Holme2011}.
Recently proposed methods of analysis for temporal networks include extensions of the methods used for static networks (\eg, distance~\cite{Kempe2000,Kostakos2009,Tang2010,Tang2010a,Tang2010b,Pan2011}, node centrality~\cite{Tang2010b,Grindrod2011,Kim2012}, community structure~\cite{Palla2007,Mucha2010,Kawadia2012}, motifs~\cite{Kovanen2011}, and components~\cite{Nicosia2012}) and specialized methods for temporal networks~\cite{Rocha2011,Lee2012,Takaguchi2011}.

In this paper, we propose an importance measure for interaction events in temporal networks.
In general, a pair of nodes may interact multiple times if the recording period is sufficiently long,
and some events may be more important than others occurring between the same node pair.
We focus on the importance of events in the sense of the amount of new information about other nodes
that can be exchanged between the pair of nodes through an event.

We develop this new method of analyzing temporal networks for two reasons.
First, the importance of nodes and links may vary over time~\cite{Kostakos2009,Tang2010,Tang2010a,Tang2010b,Grindrod2011,Kim2012}.
For example, the importance of many nodes in a social network may suddenly change when a social incident occurs~\cite{Ratkiewicz2010,Borge-Holthoefer2011}.
Professional athletes can be regarded as nodes in a directed temporal network,
and changes in the performance over the career is interpreted as the fluctuation of the node importance~\cite{Motegi2012}.
In this context, we will quantify the time-dependent importance of a link by defining an importance measure for events.

Second, from a practical standpoint, it can be easier to manipulate events rather than nodes or links for enhancing a network's performance.
A primary purpose in studying node and link centrality measures for static networks is to improve or optimize networks.
For example, it is efficient to remove appositely defined high centrality nodes to disintegrate a network and protect it from the potential spread of disease~\cite{Albert2000,Holme2002,Restrepo2008}.
Similarly, removal of high centrality links has been used to inspect the tolerance of a network against link failure~\cite{Holme2002,Motter2002,Latora2005,Onnela2007}.
If we can realize a desirable function of temporal networks
by manipulating (\eg, deleting or enhancing) a small number of single events,
it may be less costly than achieving the same outcome by manipulating the nodes or links throughout the entire period.
Anecdotally, for example, it is easier to ask a pair of individuals to stay apart for one day than to do so for the entire period.

We apply the proposed measure to real data sets
and find that event importance adequately represents the centrality of each event in the sense
that the connectivity of the remaining temporal networks drastically decreases
if we remove a small fraction of events of large importance.
We also find that event importance is broadly distributed, which implies that there is a small number of very important events
and that most events are unimportant.

\section{Methods}\label{sec:methods}
\subsection{Temporal networks}\label{sec:temporal_networks}
A temporal network~\cite{Holme2011} is defined as a series of events.
An event is composed of a particular time and pair of nodes,
which represents an interaction (\eg, conversation, email, or phone call) between the two nodes.
Although we assume that the events are undirected, extending our results to the case of directed interactions is straightforward.
The events are assumed to occur in discrete time, which reflects the time resolution of observation (\eg, 1~min).
The set of events at time $t$, where $1 \leq t \leq t_{\max}$ and $t_{\max}$ is the time of the last event in the data set,
constitutes a snapshot, that is, an unweighted network $G(t)$,
where the links connect the node pairs interacting at time $t$.
We neglect the information about the number of events between each node pair in a time unit such that there are no multiple links in $G(t)$.
If we disregard the temporal information in the data,
we can aggregate $G(t)$ into a static weighted network,
where the weight of a link is the total number of events on the link.

In temporal networks, a temporal path from node $i$ to $j$ is defined as a time-ordered event sequence satisfying the following two conditions~\cite{Kempe2000,Holme2005,Pan2011}:
(i) it begins with an event involving $i$ and ends with an event involving $j$
and (ii) one can trace a path from $i$ to $j$ by using the links in the order of the event sequence. 
For example, $i$ and $j$ are connected by a temporal path if an event between $i$ and another node $k$ occurs at time $t_1$ and an event between $k$ and $j$ occurs at time $t_2$, where $t_2 > t_1$. 
There may be no temporal path from $i$ to $j$ even if the two nodes are connected on the aggregated static network. 
 
\subsection{Vector clock and latency}\label{sec:vector_clock}
The vector clock of node $i$ is the time-dependent vector $\vect{\phi}_i (t) = \left(\phi_i^1(t), \phi_i^2(t), \ldots, \phi_i^N(t) \right)$,
where $\phi_i^\ell (t) \ (1 \leq i, \ell \leq N)$ represents the latest time among the start times of the temporal paths from $\ell$ to $i$ that terminates by time $t$~\cite{Lamport1978,Mattern1989}.
In other words, there is no temporal path that starts from node $\ell$ after time $\phi_i^\ell (t)$ and reaches node $i$ by time $t$.
The latency $b_i^\ell(t) \equiv t - \phi_i^\ell(t)$ represents the age of node $i$'s latest information about node $\ell$ at time $t$.
In general, $b_i^\ell(t) \neq b_\ell^i(t)$.
For a given event sequence, we can calculate $\vect{\phi}_i(t)$ with an efficient algorithm~\cite{Lamport1978,Mattern1989}.
The algorithm reads the events one by one in chronological order and updates $\vect{\phi}_i(t)$ of the two nodes involved in each event.
 
Our $b_i^\ell(t)$ is defined backward in that it is based on the events that occurred before time $t$.
In contrast, the authors of \cite{Pan2011} used the forward version of the temporal path length;
their definition was based on the events that occurred after time $t$.
Although the two definitions are different,
the time average of the temporal path length for any given nodes $i$ and $\ell$ is equal in the two definitions.
  
\subsection{Importance of event}\label{sec:importance}
On the basis of the vector clock (\SEC\ref{sec:vector_clock}),
Kossinets and colleagues defined the advance of event~\cite{Kossinets2008}.
The advance for node $i$ caused by an event with node $j$ at time $t$, denoted by $a_i^j(t)$,
is given by 
\begin{equation}
a_i^j(t) = \sum_{\ell \neq i} \left( \phi_i^\ell(t) - \phi_i^\ell(t-1) \right),
\label{eq:original_advance}
\end{equation}
which represents the updated amount of the latest information about other nodes (\ie, $\ell$) summed over $\ell$.
It should be noted that the right-hand side of equation~\eqref{eq:original_advance} implicitly depends on $j$;
$\phi_i^\ell(t) - \phi_i^\ell(t-1)$ is positive if and only if the event involving $i$ and $j$ at time $t$ conveys updated information about $\ell$ to $i$.

We generalize $b_i^j(t)$ and $a_i^j(t)$ to the case in which a node can be involved in multiple events in a single time unit as follows.
We set $\phi_i^j(0) = -\infty$ for all $i \neq j$ and $\phi_i^i(0) = 0$ for all $i$.
For every node $i$, we recursively define
\begin{equation}
\phi_i^\ell(t) = \max_{j \in {\cal N}_i^h} \left[  \phi_j^\ell(t-) \right] \ (1 \leq \ell \leq N),
\label{eq:update_clock}
\end{equation}
where ${\cal N}_i^h$ is the set of nodes whose distance to node $i$ in $G(t)$ is at most $h$ and
\begin{eqnarray}
\phi_j^\ell(t-) =
\left\{
\begin{array}{ll}
t  & (\ell = j),\\
\phi_j^\ell(t-1) & (1 \leq \ell \leq N, \ \ell \neq j).
\end{array}
\right.
\label{eq:update_clock_2}
\end{eqnarray}
When node $i$ is not involved in any event at time $t$, we set ${\cal N}_i^h = \left\{ i \right\}$
such that equations~\eqref{eq:update_clock} and \eqref{eq:update_clock_2} imply $\phi_i^\ell (t) = \phi_i^\ell(t-1)$ $(\ell \neq i)$.
Then, we set $b_i^\ell(t) = t - \phi_i^\ell(t)$ $(1 \leq \ell \leq N)$ as before. 
The positive integer $h$, called the horizon in \cite{Tang2010,Tang2010a}, specifies the range of information spreading in a time unit.

When node $i$ is involved in multiple events in a single time unit,
we determine the contribution of each neighbor to the advance of the information about others as follows.
First, for given nodes $i$ and $\ell$, we identify the nodes in ${\cal N}_i^h$ that give the maximum value of the right-hand side of equation~\eqref{eq:update_clock}.
Such nodes have the latest information about node $\ell$ among the nodes in ${\cal N}_i^h$.
Second, we determine the so-called contributing neighbors of node $i$.
It is defined as node $i$'s neighbors on $G(t)$ such that they are on a shortest path between $i$ and a node in ${\cal N}_i^h$ having the latest information about node $\ell$.

We assume that all contributing neighbors equally contribute to information passing from node $\ell$ to node $i$.
We define the advance of node $i$ by contributing neighbor $j$ by
\begin{equation}
a_i^j(t) = \sum_{\ell: M_i(\ell;t) \ni j} \frac{\Delta\phi_i^\ell(t)}{\left| M_i(\ell;t) \right|}, 
\label{eq:advance}
\end{equation}
where $M_i(\ell;t)$ is the set of $i$'s contributing neighbors with regard to the information about node $\ell$, and
\begin{eqnarray}
\Delta\phi_i^\ell(t) =
\left\{
\begin{array}{ll}
0 & ( \phi_i^\ell(t-1) = -\infty \ {\rm and} \ \phi_i^\ell(t) = -\infty),\\
\phi_i^\ell(t) & ( \phi_i^\ell(t-1) = -\infty \ {\rm and} \ \phi_i^\ell(t) \neq -\infty),\\
\phi_i^\ell(t) - \phi_i^\ell(t-1) & ({\rm otherwise}).
\end{array}
\right.
\label{eq:deltaphi_i}
\end{eqnarray}
In the first case on the right-hand side of equation~\eqref{eq:deltaphi_i},
there is no temporal path from node $\ell$ to node $i$ by time $t$.
In the second case, the temporal path from node $\ell$ reaches node $i$ for the first time at $t$.
In the third case, the temporal path from node $\ell$ to node $i$ established before $t$ is renewed at $t$.

\begin{figure}
\centering
\includegraphics[width=0.4\hsize, clip]{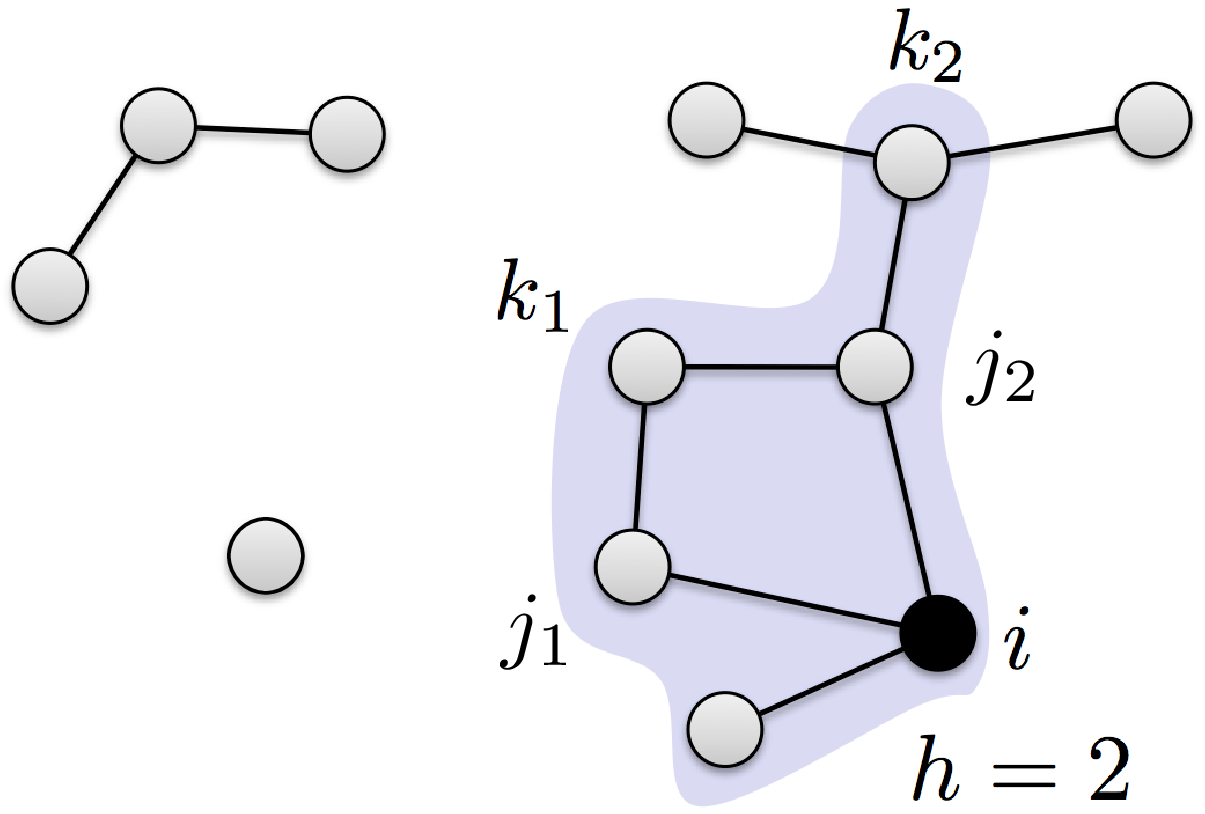}
\caption{Schematic of a snapshot of a temporal network.}
\label{fig:advance_schematic}
\end{figure}

For expository purposes, we set $h=2$ and focus on node $i$ in the snapshot $G(t)$ shown in \FIG\ref{fig:advance_schematic}.
First, if $k_1$ is the only node in ${\cal N}_i^2$ that has the latest information about $\ell$ at $t$,
$j_1$ and $j_2$ are $i$'s contributing neighbors.
Therefore, $M_i(\ell; t) = \left\{ j_1, j_2 \right\}$
such that $j_1$ and $j_2$ contribute $\Delta\phi_i^{\ell}(t) / 2$ to $a_i^{j_1}(t)$ and $a_i^{j_2}(t)$, respectively.
Second, if multiple nodes in ${\cal N}_i^h$ have the latest information about $\ell$ at $t$,
we assume that only the nodes closest to $i$ convey the information.
If $j_2$ and $k_1$, for example, have the latest information about $\ell$ at $t$,
$j_2$ but not $k_1$ conveys the information such that $j_2$ is node $i$'s sole contributing neighbor.
Therefore,  $M_i(\ell; t) = \left\{ j_2 \right\}$ and $j_2$ contributes $\Delta\phi_i^{\ell}(t)$ to  $a_i^{j_2}(t)$,
and $a_i^{j_1}(t)$ does not change.
Third, suppose that only $k_1$ and $k_2$ have the latest information about $\ell$ at $t$.
Then, we assume that $j_1$ and $j_2$ are $i$'s contributing neighbors and that the two nodes contribute equally to the advance,
although $j_1$ is on the shortest path from $k_1$ to $i$, whereas $j_2$ is on the shortest paths from both $k_1$ and $k_2$ to $i$.
Therefore, $M_i(\ell; t) = \left\{ j_1, j_2 \right\}$ such that $j_1$ and $j_2$ contribute $\Delta\phi_i^{\ell}(t) / 2$ to $a_i^{j_1}(t)$ and $a_i^{j_2}(t)$, respectively.
 
Even though the events are defined as undirected, $a_i^j(t) \neq a_j^i(t)$ in general.  
We define the importance of the event between nodes $i$ and $j$ at time $t$ by
\begin{equation}
I_{ij}(t) = \frac{a_i^j(t) + a_j^i(t)}{2}.
\label{eq:importance}
\end{equation}

\subsection{Empirical data}
We measure $I_{ij}(t)$ for three real data sets.
All the data sets were obtained from the observation of face-to-face interactions.
Basic statistics of the data sets are summarized in \TAB\ref{tab:data_info}.
Two data sets are the interaction logs between office workers in two different Japanese companies;
they were collected by World Signal Center, Hitachi, Ltd., Japan~\cite{Takaguchi2011,Yano2009,Wakisaka2009}. 
We call them Office1 and Office2, respectively, although they were called $D_1$ and $D_2$ in our previous paper~\cite{Takaguchi2011}.
The numbers of events in Office1 and Office2 are larger than those in $D_1$ and $D_2$,
because we merged repeated events between the same node pair in consecutive time bins into one event for $D_1$ and $D_2$~\cite{Takaguchi2011}
but did not do so for the Office1 and Office2 data sets.  
The third data set (called Conference) is the interaction record between the attendees at a scientific conference, collected by the SocioPatterns collaboration~\cite{Isella2011}.

\section{Results}
In this section, we set $h=N-1$, which corresponds to the situation in which information instantaneously spreads to all nodes in each connected component in a snapshot.
For Office1 data set, we also confirmed that the results are qualitatively the same in the other extreme case $h=1$,
in which the information is propagated by only one hop in a single time unit (supplementary figure~S1 in supplementary data).

\subsection{Heterogeneity in the importance of events}
\begin{figure}
\centering
\includegraphics[width=0.4\hsize, clip]{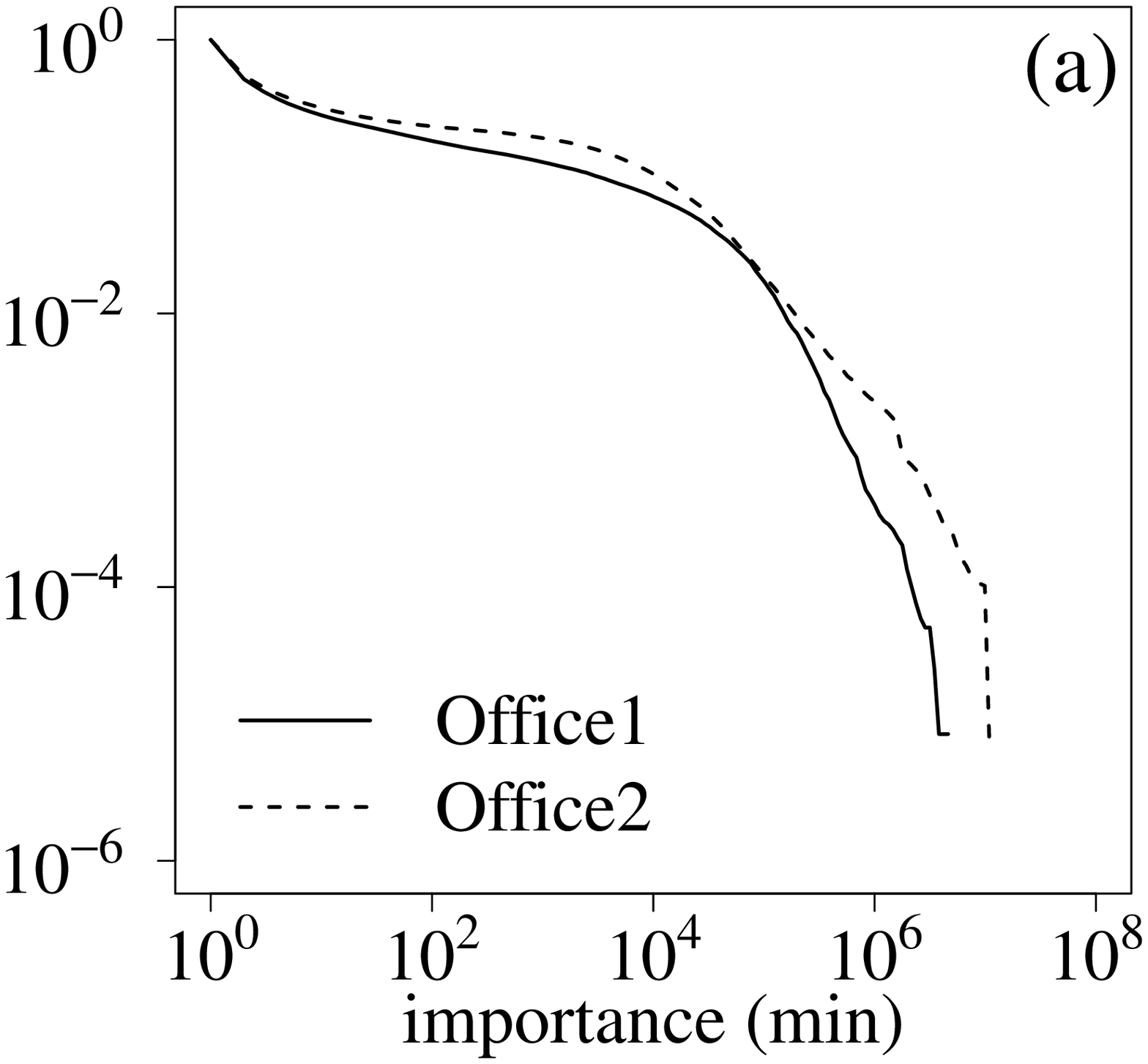}
\includegraphics[width=0.4\hsize, clip]{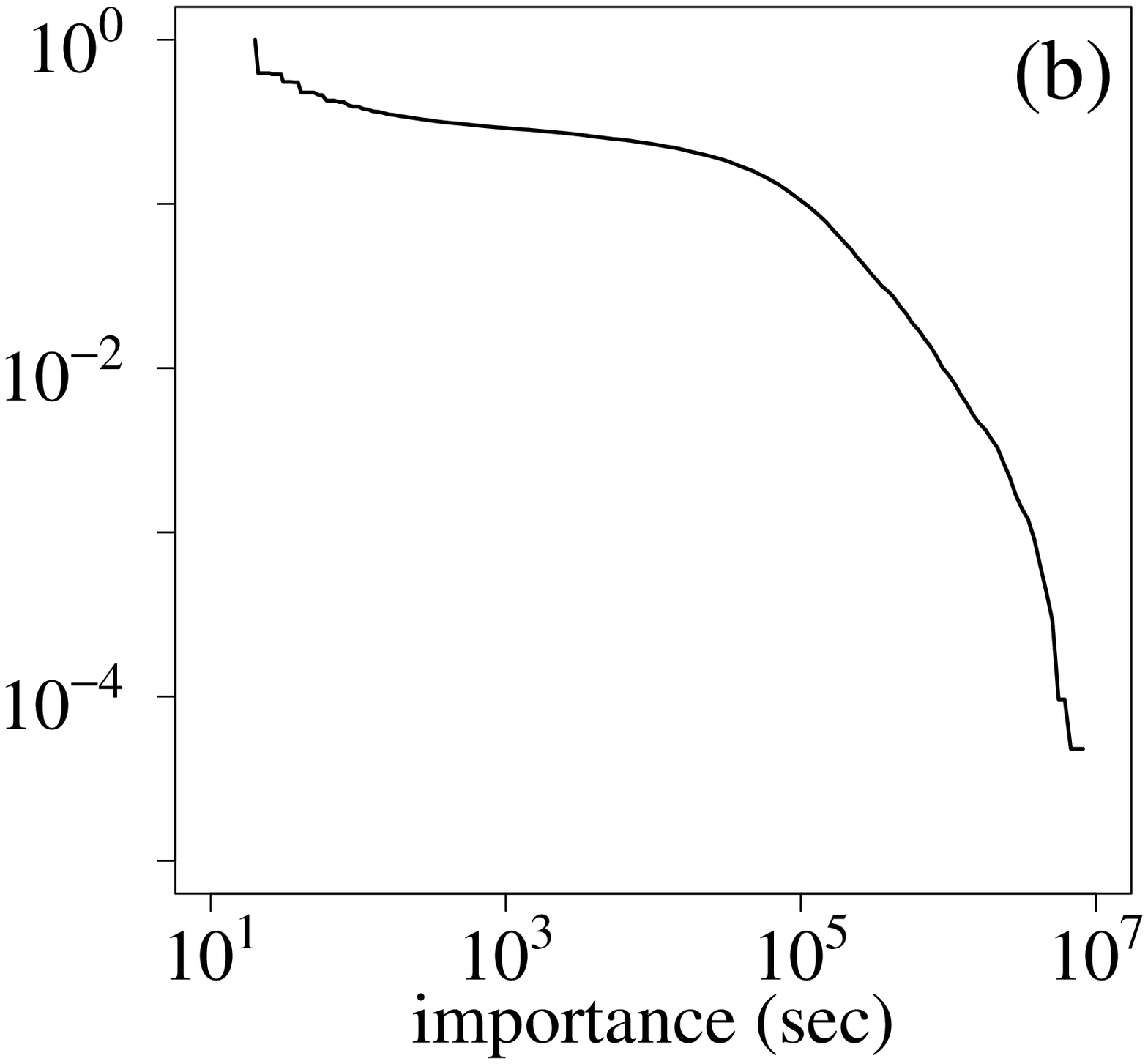}
\caption{Complementary cumulative distribution of the importance of event.
(a) Office1 and Office2 data sets. (b) Conference data set.}
\label{fig:hist_importance}
\end{figure}

The complementary cumulative distributions of $I_{ij}(t)$ (\ie, $\Prob \left( I_{ij}(t) \geq I\right)$)
for the three data sets are shown in \FIG\ref{fig:hist_importance}.
The $I_{ij}(t)$ values are broadly distributed for all the data sets,
which implies that a small fraction of events has large importance values and most events have small importance values.

The advance of event is strongly asymmetric.
For Office1 and Conference data sets, the frequency of events having specified $\max\left[ a_i^j(t), a_j^i(t)\right]$
and $\min\left[ a_i^j(t), a_j^i(t)\right]$ values are shown in \FIG\ref{fig:asym_advance} (see supplementary figure~S2(a) for Office2 data set).
If the advance were symmetric,  that is $a_i^j(t) = a_j^i(t)$, the frequency would be concentrated on the diagonal.
However, \FIG\ref{fig:asym_advance} suggests that most events have very different values of $a_i^j(t)$ and  $a_j^i(t)$.
 
\subsection{Event removal tests}\label{sec:event_removal_tests}
To examine if $I_{ij}(t)$ represents the importance of events in bridging temporal paths,
we investigate the connectivity of the temporal networks after we remove a fraction of events.
The procedure of an event removal test is similar to that of link removal in static networks~\cite{Holme2002,Motter2002,Latora2005,Onnela2007}.
We remove events according to the (i) ascending order of the importance, (ii) descending order of the importance,
(iii) ascending order of the link weight (\ie, the total number of events between the node pair), (iv) descending order of the link weight, and (v) random order.
In schemes (i) and (ii), we do not recalculate $I_{ij}(t)$ after removing each event.
In schemes (iii) and (iv), we remove a randomly selected event on the link with the smallest and largest link weight, respectively,
in each time step.
Then, we recalculate the weight of the link from which an event is removed and repeat the removal procedure.
If $I_{ij}(t)$ is an adequate measure of the importance of events,
the connectivity of temporal networks would decrease more upon the removal of a specified number of events with large $I_{ij}(t)$ (\ie, scheme (ii)) than with small $I_{ij}(t)$ (\ie, scheme (i)).
  
We use two quantities to measure the connectivity of the remaining temporal networks.
First, we define the reachability ratio $f$ by the fraction of ordered pairs $(i,j)$ $(i \neq j)$ such that there is at least one temporal path from $i$ to $j$~\cite{Holme2005}.
Second, we define the network efficiency $E$~\cite{Tang2010} by
\begin{equation}
E = \frac{1}{N(N-1)} \sum_{(i,j), \ i \neq j} \frac{1}{\overline{b}_i^j},
\end{equation}
where $\overline{b}_i^j$ denotes the time average of $b_i^j(t)$.
A problem with time averaging is that $b_i^j(t)$ is indefinite
until the first temporal path from $j$ reaches $i$.
We address this problem by virtually replicating the last temporal path between each node pair immediately before $t=0$.
This boundary condition is a variant of that proposed in a recent study~\cite{Pan2011}.   
When no node pair is connected by a temporal path,
we obtain $\overline{b}_i^j = \infty$ for any $1 \leq i, j \leq N$.
In this case,  $E$ takes the minimum value of zero.
$E$ is positive but small when many pairs of nodes are connected via long temporal paths (\ie, a large $\overline{b}_i^j$).
In contrast, $E$ is large when many pairs of nodes are connected via short temporal paths.
We use the two measures because $f$ is more intuitive than $E$ and $E$ is finer than $f$.

The dependence of $f$ and $E$ on the fraction of removed events is shown in \FIG\ref{fig:original_event_removal} for Office1 and Conference data sets.
The values of $f$ and $E$ are normalized by the values in the case of no event removal in this and all of the following figures. 
The results for the two data sets are similar.
We also confirmed that Office2 data set also yields similar results (supplementary figures~S2(b) and S2(c)).
Figure~\ref{fig:original_event_removal} indicates that removing 80\% of the events in the ascending order of $I_{ij}(t)$ has little effect on $f$ and $E$
and that removing 20\% of the events in the descending order of $I_{ij}(t)$ drastically decreases $f$ and $E$.
Therefore, $I_{ij}(t)$ adequately represents the importance of event
in the sense that a small fraction of events with large $I_{ij}(t)$ values plays a crucial role in sustaining temporal paths.

Of the five removal schemes, event removal in the ascending order of the link weight yields
the largest decrease of $f$ and $E$ at a small fraction of removed events for both data sets.
This result is derived from the so-called ``strength of weak ties'' property~\cite{Granovetter1973}
of the aggregated networks corresponding to these two temporal networks.
The strength of weak ties property claims that
weak links (\ie, links with small weights) bridge the communities (\ie, dense subgraphs) that mainly contain strong links (\ie, links with large weights).
We confirmed this property for Office1 and Office2 data sets in our previous study~\cite{Takaguchi2011}.
We found that the same property holds true for Conference data set (Appendix~A).
Removing events on weak links tends to fragment the aggregated network into disconnected components
such that temporal paths between any two nodes in different components are lost.
 
A possible criticism is that it is not necessary to use $I_{ij}(t)$ when evaluating the importance of events,
because removing events on weak links most efficiently makes the temporal network disconnected.
However, $I_{ij}(t)$ seems to be a better measure
because events with large $I_{ij}(t)$ are necessary and sufficient for sustaining temporal paths.
The removal of a small fraction (\eg, 20\%) of events with the largest $I_{ij}(t)$ drastically reduces $E$,
and the same set of events sustains the temporal paths
such that the values of $f$ and $E$ are almost the same as those for the original temporal network.     
In contrast, removing a small fraction of events on weak links admittedly fragments the networks as shown in \FIG\ref{fig:original_event_removal}.
However, the same set of events does not sustain efficient temporal paths; 40\% of events on weak links are needed to recover efficient temporal paths (Figs.~\ref{fig:original_event_removal}(b) and \ref{fig:original_event_removal}(d)).
We also confirmed that the Spearman's rank correlation between $I_{ij}(t)$ averaged over all the events on a link and the link weight is only weakly negative; the coefficient values are equal to $-0.4078$, $-0.2370$, and $-0.3891$ for Office1, Office2, and Conference data sets, respectively.
 
Proxy quantities to $I_{ij}(t)$ other than that based on weak links may exist.
An event that occurs after a long interevent interval (IEI) since the last event between the same node pair is expected to have large $I_{ij}(t)$,
because $i$ $(j)$ has not obtained up-to-date information that $j$ $(i)$ may have about itself and others for a long time.
We calculate the Spearman's rank correlation coefficient between $I_{ij}(t)$ and (i) the length of the IEI since the last event between the node pair, (ii) the number of events in the entire temporal network within the last IEI, (iii) the number of events involving either $i$ or $j$ within the last IEI, and (iv) the number of nodes that interact with $i$ or $j$ within the last IEI.
The correlation coefficients for Office1 data set
are equal to $0.819$, $0.701$, $0.701$, and $0.631$, for cases (i), (ii), (iii), and (iv), respectively.
The length of the last IEI approximates $I_{ij}(t)$ most accurately among the four.
The results for the event removal test based on the order of the last IEI are similar to those for the event removal based on $I_{ij}(t)$ (supplementary figure~S3).  

\subsection{Event removal tests for randomized temporal networks}\label{sec:origin}
\begin{figure}
\centering
\includegraphics[width=0.4\hsize, clip]{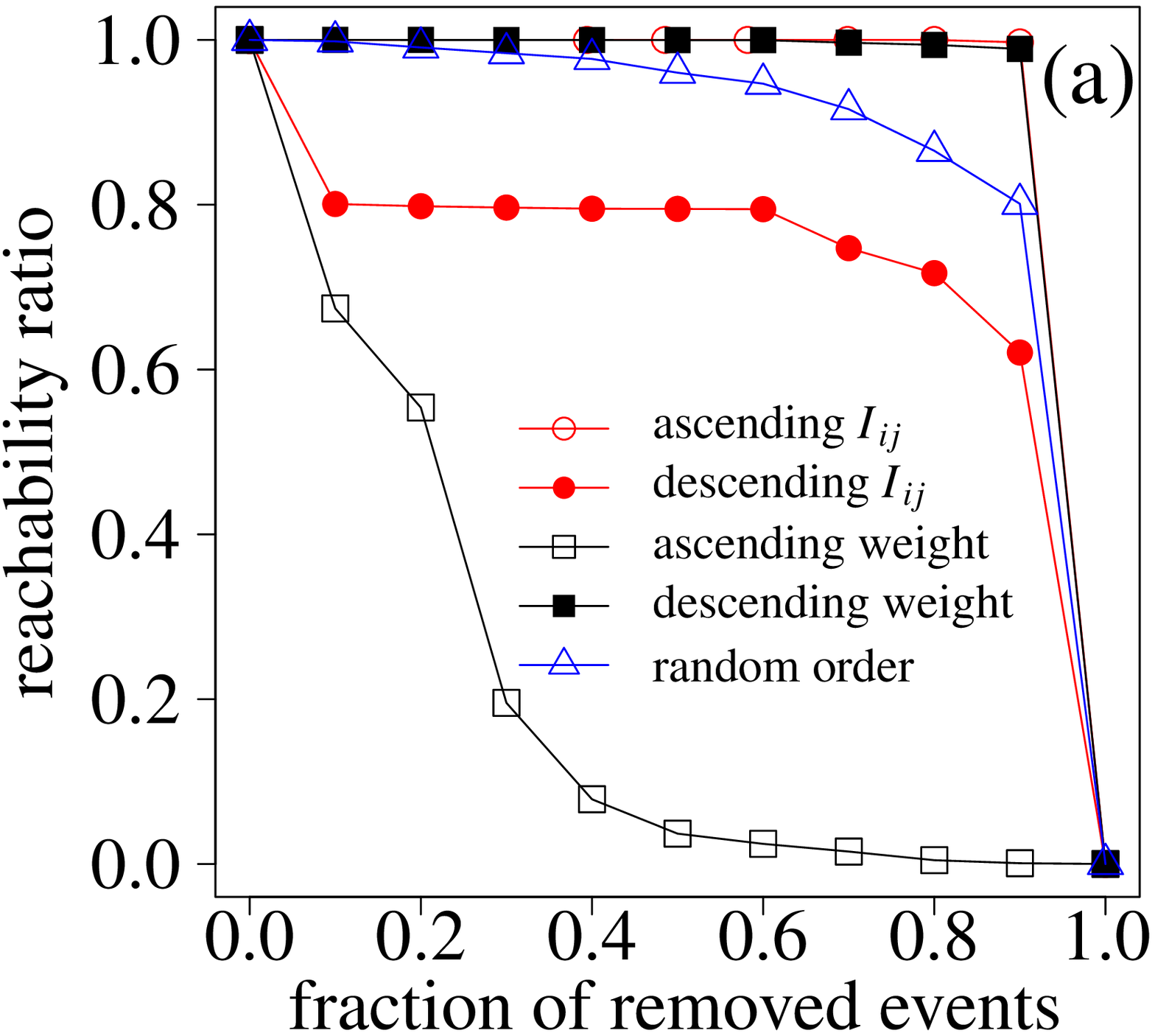}
\includegraphics[width=0.4\hsize, clip]{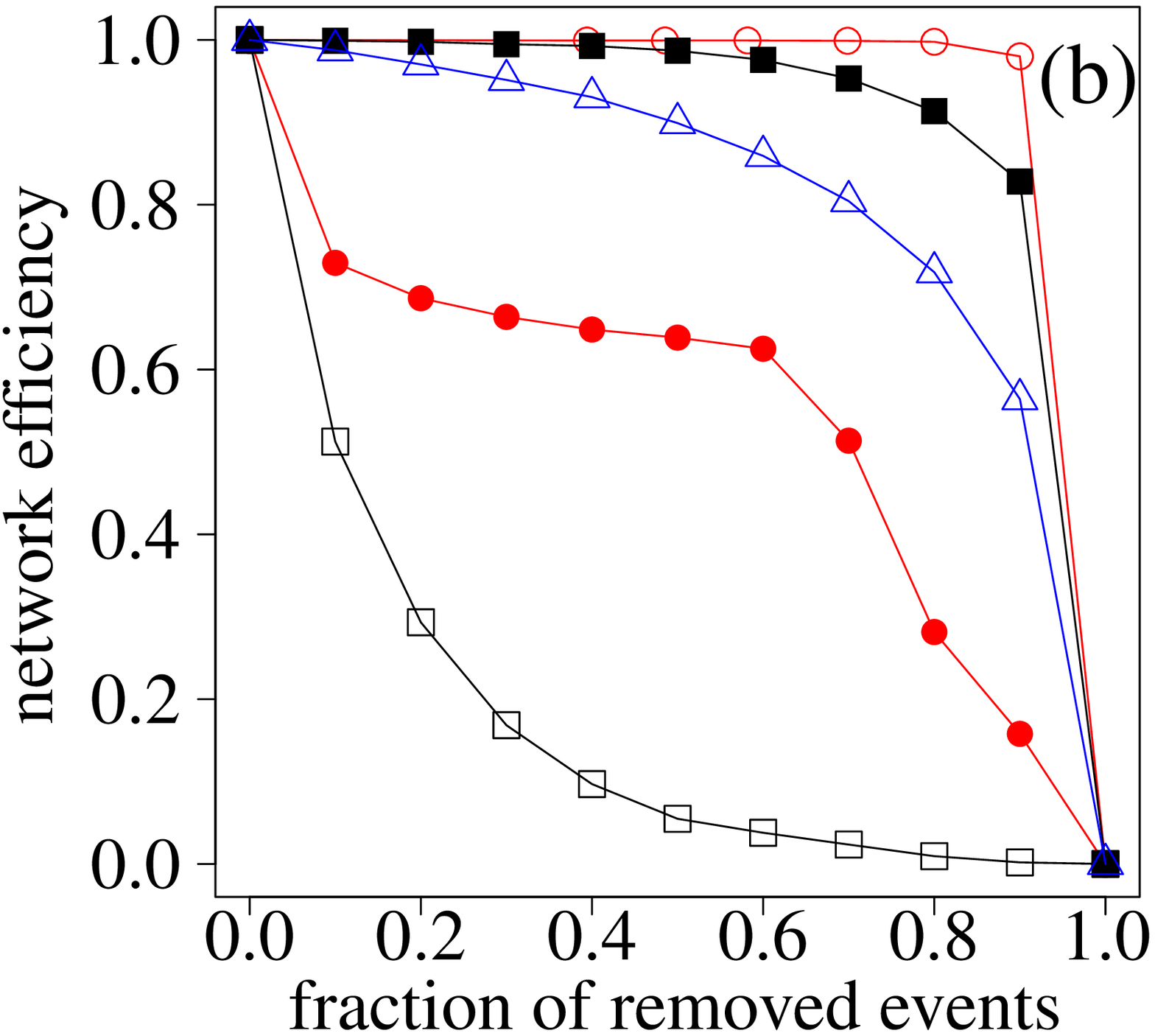}\\
\includegraphics[width=0.4\hsize, clip]{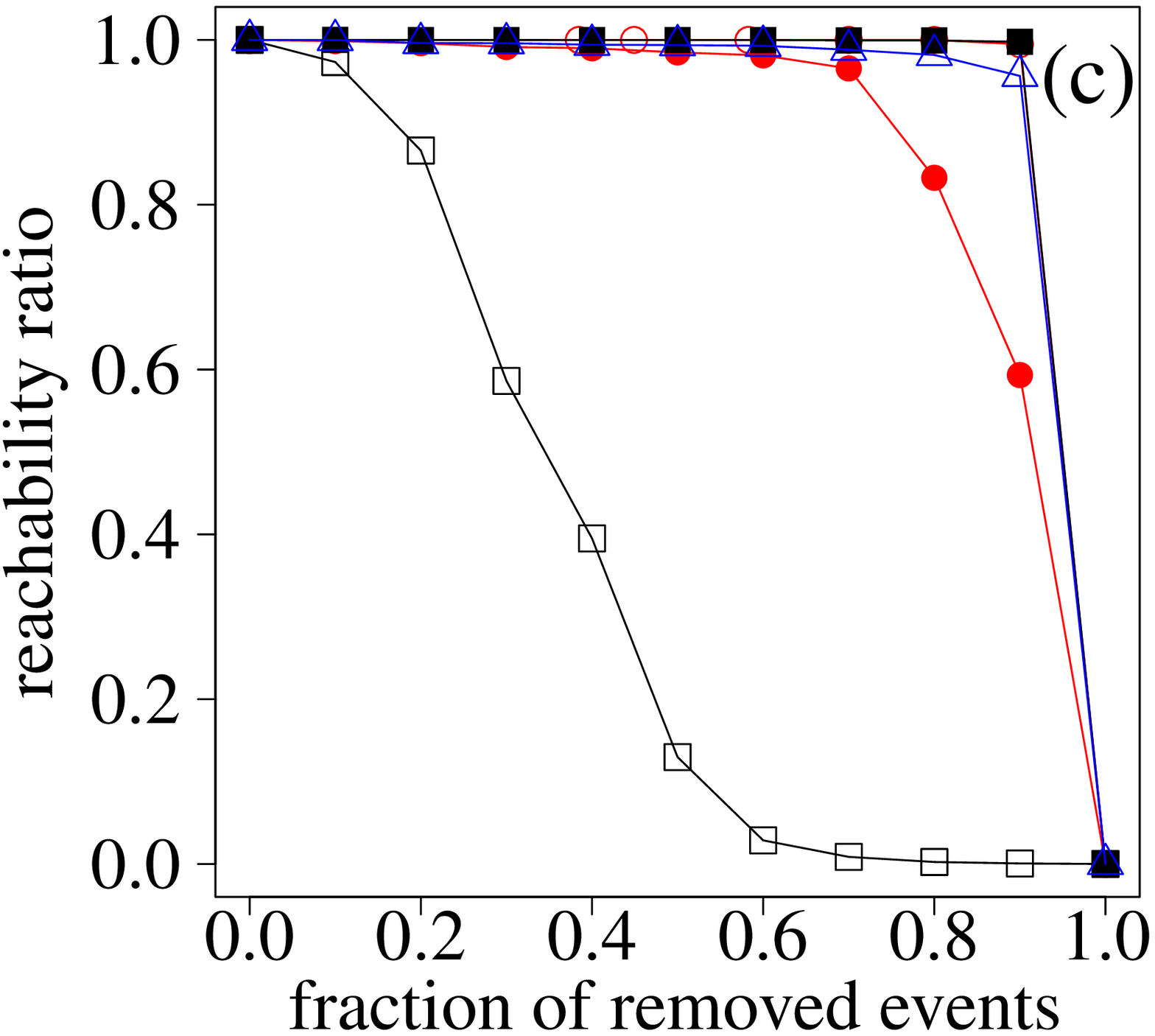}
\includegraphics[width=0.4\hsize, clip]{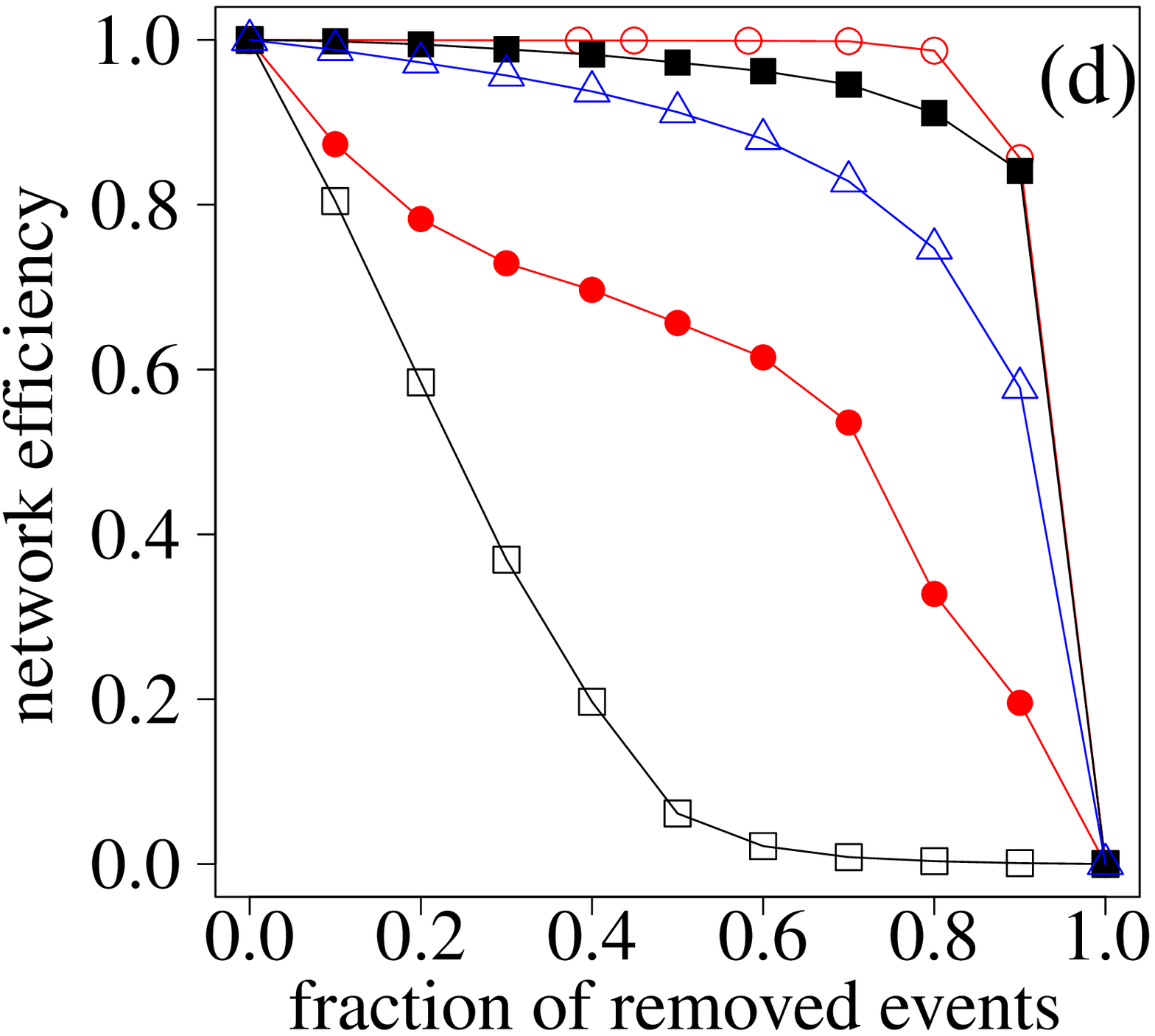}
\caption{Results of the event removal tests. (a), (c) Reachability ratio (\ie, $f$).
(b), (d) Network efficiency (\ie, $E$).
(a), (b) Office1 and (c), (d) Conference data sets.
}
\label{fig:original_event_removal}
\end{figure}

We showed that a small fraction of events with the largest $I_{ij}(t)$ can sustain efficient temporal paths (\FIG\ref{fig:original_event_removal}),
which we call the robustness property.
In this section, we seek the origins of the robustness property by carrying out event removal tests for randomized temporal networks.

We randomize the original temporal networks in two ways.
First, we randomly shuffle the IEIs for each link while keeping the times of the first and last events.
This shuffling conserves the distribution of the IEI and the structure of the aggregated network
and eliminates all other temporal structure of the IEIs.
Second, we generate so-called Poissonized IEIs by reassigning to each event a random event time
that is distributed uniformly and independently on $[0,t_{\max}]$,
where $t_{\max}$ is the time of the last event in the original temporal network.
The event sequence on each link then independently obeys the Poisson process
such that the temporal structure of the IEIs including the IEI distribution is destroyed,
although the aggregated network is unaffected.
Properties of temporal network that are conserved and dismissed as a result of different randomization methods are summarized in table~\ref{tab:randomization}.

For Office1 data set, the $E$ values for the temporal networks generated by the shuffled IEIs and the Poissonized IEIs
are shown in Figs.~\ref{fig:E_RI_RT_RG_event_removal}(a) and \ref{fig:E_RI_RT_RG_event_removal}(b), respectively
(see supplementary figure~S4 for $f$ for the same data).
We obtain qualitatively the same results for Office2 and Conference data sets (supplementary figures~S5 and S6).
The results for the shuffled IEIs (\FIG\ref{fig:E_RI_RT_RG_event_removal}(a)) are qualitatively the same as
those for the original temporal network (\FIG\ref{fig:original_event_removal}(b)).
In particular, $E$ changes little when approximately 80\% of events with the smallest $I_{ij}(t)$ values are removed.
The results for the Poissonized IEIs (\FIG\ref{fig:E_RI_RT_RG_event_removal}(b)) are
considerably different from those for both the original temporal network (\FIG\ref{fig:original_event_removal}(b)) and the shuffled IEIs (\FIG\ref{fig:E_RI_RT_RG_event_removal}(a)).
With the Poissonized IEIs, $E$ decreases considerably upon the removal of a relatively small fraction of events with the smallest $I_{ij}(t)$.
Therefore, a long-tailed IEI distribution is a necessary condition for the robustness property.
As a remark, the values of $f$ for the original temporal network (\FIG\ref{fig:original_event_removal}(a)),
the shuffled IEIs (supplementary figure~S4(a)), and the Poissonized IEIs  (supplementary figure~S4(b)) are similar,
probably because $f$ is not very sensitive to the IEI distribution.

These results lead us to hypothesize that long-tailed IEI distributions rather than the structure of aggregated networks,
such as a heterogeneous degree distribution, primarily contributes to the robustness property.
Therefore, we implement a third randomization scheme in which we randomly rewire links in the aggregated network while keeping the event sequence on each link.
If the generated network is disconnected as a static network, we discard the realization and redo the rewiring.
This randomization eliminates the properties of aggregated networks, such as the heterogeneous degree distribution, community structure, and the strength of weak ties property.
The rewiring randomization conserves the IEI distribution on each link and the distribution of the link weight (table~\ref{tab:randomization}).
The results of the event removal tests for this randomization (\FIG\ref{fig:E_RI_RT_RG_event_removal}(c))
are similar to those for the original temporal network (\FIG\ref{fig:original_event_removal}(b))
and those for the shuffled IEIs (\FIG\ref{fig:E_RI_RT_RG_event_removal}(a)).
Therefore, the structure of the aggregated network has little effect on the robustness property.

The results for the two types of randomized temporal networks shown in Figs.~\ref{fig:E_RI_RT_RG_event_removal}(a) and \ref{fig:E_RI_RT_RG_event_removal}(c) are similar to those for the original temporal network (\FIG\ref{fig:original_event_removal}(b)),
but both types of randomization simultaneously conserve the long-tailed IEI distribution and the distribution of the link weight.
To investigate the sole contribution of the long-tailed IEI distribution,
we carry out the following event removal tests for temporal networks generated as follows.
We first generate a regular random graph having $N=163$ nodes, the same as Office1,
and the degree of each node $26$ which is close to the average degree of Office1.
Then, we place an event sequence on each link such that the IEIs on each link are independently drawn from the distribution $p(\tau)$.
We set the number of events on each link to $60$, which is also similar to the average for Office1. 
The precise procedure for generating the temporal networks is described in Appendix~B.
The aggregated network of the generated temporal network is devoid of a heterogenous distribution of link weight.

\begin{figure}
\centering
\includegraphics[width=0.4\hsize,clip]{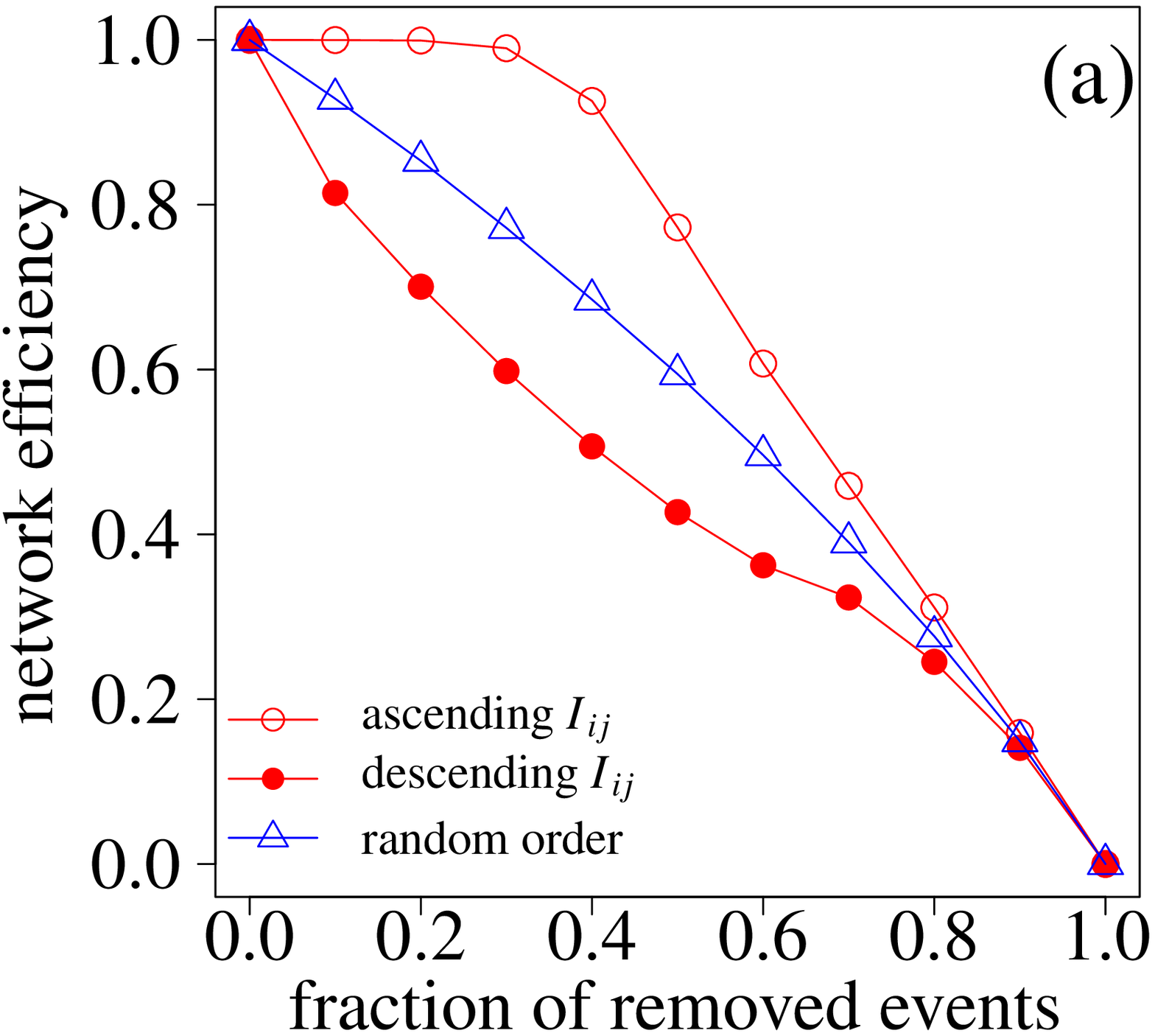}
\includegraphics[width=0.4\hsize,clip]{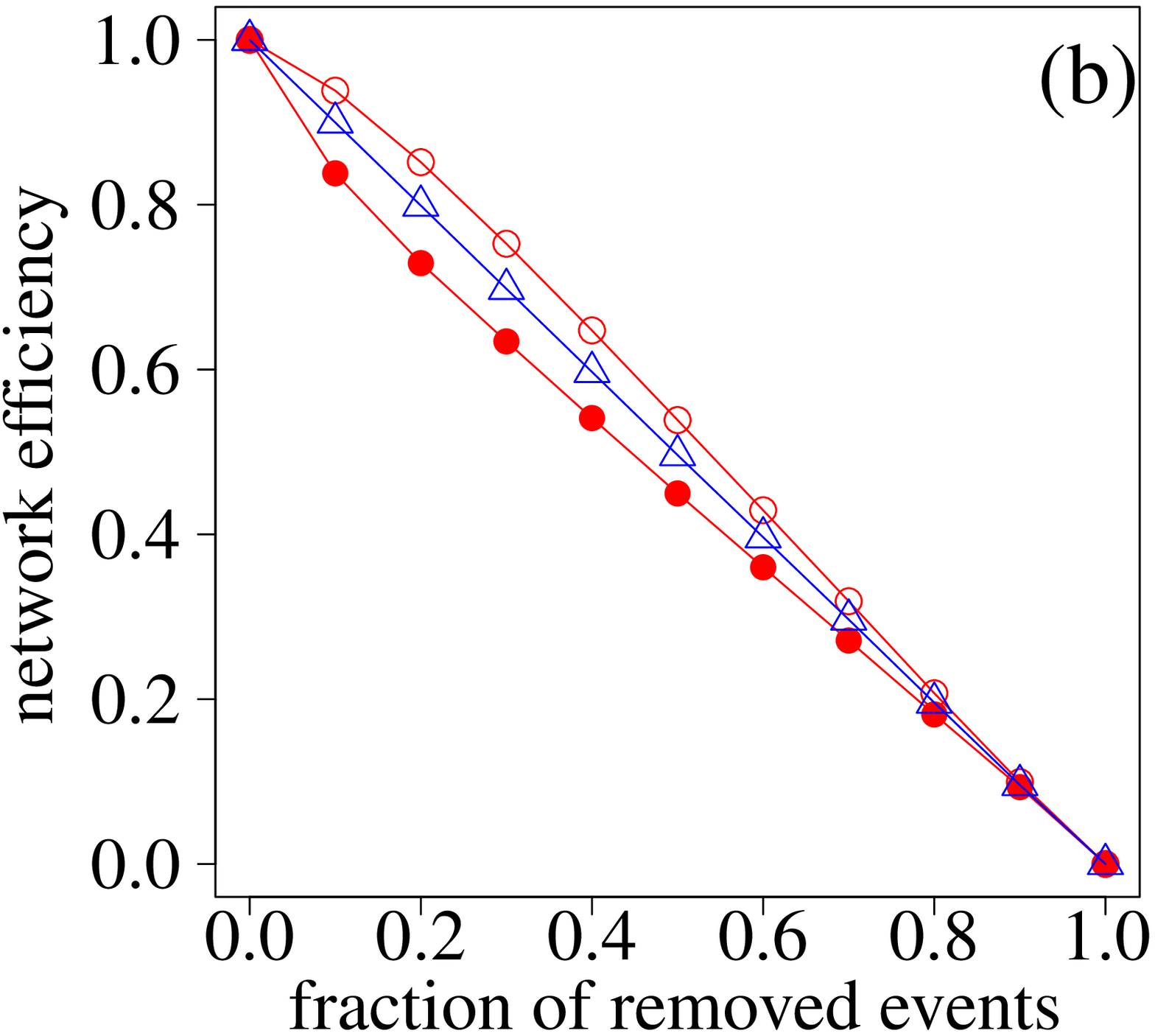}
\caption{Network efficiency for the temporal networks built on the regular random graph.
(a) Long-tailed and (b) exponential IEI distributions.
}
\label{fig:RRG_event_removal}
\end{figure}

In \FIG\ref{fig:RRG_event_removal}(a), we plot $E$ for the long-tailed IEI distribution, that is, $p(\tau) \propto \tau^{-1}\exp(-\tau/1000)$,
mimicking statistics observed in human communication behavior~\cite{AVazquez2007,Candia2008}.
With the long-tailed $p(\tau)$,
the value of $E$ changes little upon the removal of the 30\% of events with the smallest $I_{ij}(t)$ values.
In contrast, for the exponential IEI distribution, $E$ decreases upon even a small fraction of removed events irrespectively of the scheme of event removal, that is, ascending or descending order of the importance and random order (\FIG\ref{fig:RRG_event_removal}(b)).
Therefore, bursty activity patterns explain the robustness property to a large extent.
The remaining contribution may be explained by other factors, including the heterogeneity in the link weight.

\section{Discussion}
We proposed a centrality measure for interaction events in temporal networks.
An important event is defined as one that conveys a large amount of new information 
to the two individuals involved in the event.
Our main finding is the robustness property of temporal networks such
that the connectivity of temporal networks remains almost the same after a large fraction of events with small importance values is removed.
Conversely, connectivity is destroyed after a small fraction of events with large importance values is removed.
We also found that the importance of an event is broadly distributed
and that the advance of an event is strongly asymmetric for the two nodes involved in the event.
Bursty nature of interaction events, not the structural properties of the aggregated networks including the heterogeneous degree distribution,
is a main contributor to the robustness property.
 
Although our results suggest that events with small importance values are unnecessary for efficient communication,
such redundant events may be practically necessary.
For example, two individuals may need repeated interactions within a short interval for the purpose of persuasion or negotiation.
This is an obvious and important limit of the present study.
To cope with this issue, we need additional information about interactions
such as the contents of conversations and status of individuals in an organization.
Nevertheless, we hope that the present framework serves to improve our understanding of the meaning of each event in temporal networks. 

We symmetrized $a_i^j(t)$ to define the importance of an event in equation~\eqref{eq:importance}.
However, the asymmetry in $a_i^j(t)$ is expected to contain rich information about directed relationships between individuals. 
For example, assume that individual X tends to have new information about many others, perhaps through frequent events with others.
X may give more up-to-date information about others to neighbor Y in each event than X receives up-to-date information from Y.
In this case, X may be more important than Y in this dyadic relationship.
It should be noted that the original temporal and aggregated networks are symmetric,
and it may be useful to analyze the static directed network constructed by aggregating but not symmetrizing the $a_i^j(t)$ values on each link
to reveal key individuals and information propagation on temporal networks.
Analytical tools to this end include those specialized for directed networks, such as the PageRank,
network motifs, and reciprocity~\cite{Newman2010book}. 

In general, temporal information may be useful for preventing epidemics in temporal networks~\cite{Holme2011,Lee2012}.
The concept of the importance of an event may be useful for this purpose.
A node $i$ involved in an event with large $a_i^j(t)$ gains short temporal paths from other nodes.
A short temporal path may serve as an efficient pathway of epidemic spreading.
If an important event occurs,
potential events on the same link occurring immediately after this trigger event may also efficiently propagate epidemics, although such successive events carry short IEIs and therefore are likely to have small importance values.
Then, an effective prevention method may be to prohibit the occurrence of successive events once an event with a large importance value is detected.
Fortunately, we defined the importance of an event based on the events in the past only
and did not require the information about the events in the future.
Therefore, we can implement such a prevention method as an online algorithm.
Although the proposed prevention method is an intervention on links,
the importance of the link in this sense generally fluctuates over time.
This type of nonstationarity may be induced by external shocks to the temporal network.

\section*{Acknowledgments}
The authors thank to the SocioPatterns collaboration (http://www.sociopatterns.org) for providing the data set.
TT acknowledges the support provided through Grants-in-Aid for Scientific Research (No.~10J06281) from JSPS, Japan.
NM acknowledges the support provided through Grants-in-Aid for Scientific Research (No.~23681033, and Innovative Areas ``Systems Molecular Ethology'' (No.~20115009)) from MEXT, Japan.

\section*{Appendix}
\subsection*{A. Strength of weak ties property in Conference data set}\label{sec:weakties_conference}
We previously found the strength of weak ties property in Office1 ($D_1$) and Office2 ($D_2$) data sets~\cite{Takaguchi2011}
by showing a positive correlation between the neighborhood overlap~\cite{Onnela2007} and the link weight,
which is the number of events between the node pair in the entire recording period.
The neighborhood overlap of link $(i,j)$, denoted by $O_{ij}$, is defined by
\begin{equation}
O_{ij} = \frac{\left| {\cal N}_i \cap {\cal N}_j \right|}{\left| {\cal N}_i \cup {\cal N}_j \right|-2},
\end{equation}
where ${\cal N}_i$ is the set of node $i$'s neighbors in the aggregated network,
and $\left| \cdot \right|$ is the number of elements in the set.
$O_{ij}$ takes the minimum value zero when nodes $i$ and $j$ do not share any neighbor
and the maximum value unity when nodes $i$ and $j$ share all the neighbors.
If a network has the strength of weak ties property,
links with large (small) weights tend to connect intracommunity (intercommunity) node pairs and hence have large (small) $O_{ij}$.
In \FIG\ref{fig:Oij_conference}, for Conference data set, $O_{ij}$ averaged over the links with $w_{ij} < w$, denoted by $\langle O \rangle_w$, is plotted against the fraction of links with $w_{ij} < w$, denoted by $P_{\rm cum}(w)$.
Because  $\langle O \rangle_w$ monotonically increases with $P_{\rm cum}(w)$, the aggregated network of Conference data set has the strength of weak ties property.

\subsection*{B. Temporal networks on the regular random graphs}
In this section, we describe the procedure for generating the temporal networks on a regular random graph.
A similar algorithm for generating temporal networks based on IEIs on nodes, not links, was recently proposed~\cite{Rocha2012}.
First, we generate a regular random graph with $N=163$ nodes and degree $26$ by using the configuration model~\cite{Newman2001,Newman2010book}.
Second, we generate the so-called template IEI sequence composed of $60-1=59$ IEIs whose length independently obeys a long-tailed distribution given by $p(\tau) \propto \tau^{-1} \exp(-\tau/1000)$~\cite{Clauset2009}.
Third, we assign to each link an initial event time $t_0$ and a sequence of the IEIs generated by randomly shuffling the template IEI sequence.
For each link, $t_0$ is independently drawn from the uniform distribution on $[0, 100]$.
Each link has the same number of events in the generated network.

To generate the temporal network with the exponential IEI distribution on the regular random graph,
we randomized the event times in the temporal network with the long-tailed IEI distribution.
In other words, we generate a temporal network according to the procedure described above
and reassign to each event a random event time that is distributed uniformly and independently on $[0,t_{\max}]$,
where $t_{\max}$ is the time of the last event in the entire temporal network.


\clearpage

\clearpage

\clearpage
\begin{figure}
\centering
\includegraphics[width=0.403\hsize, clip]{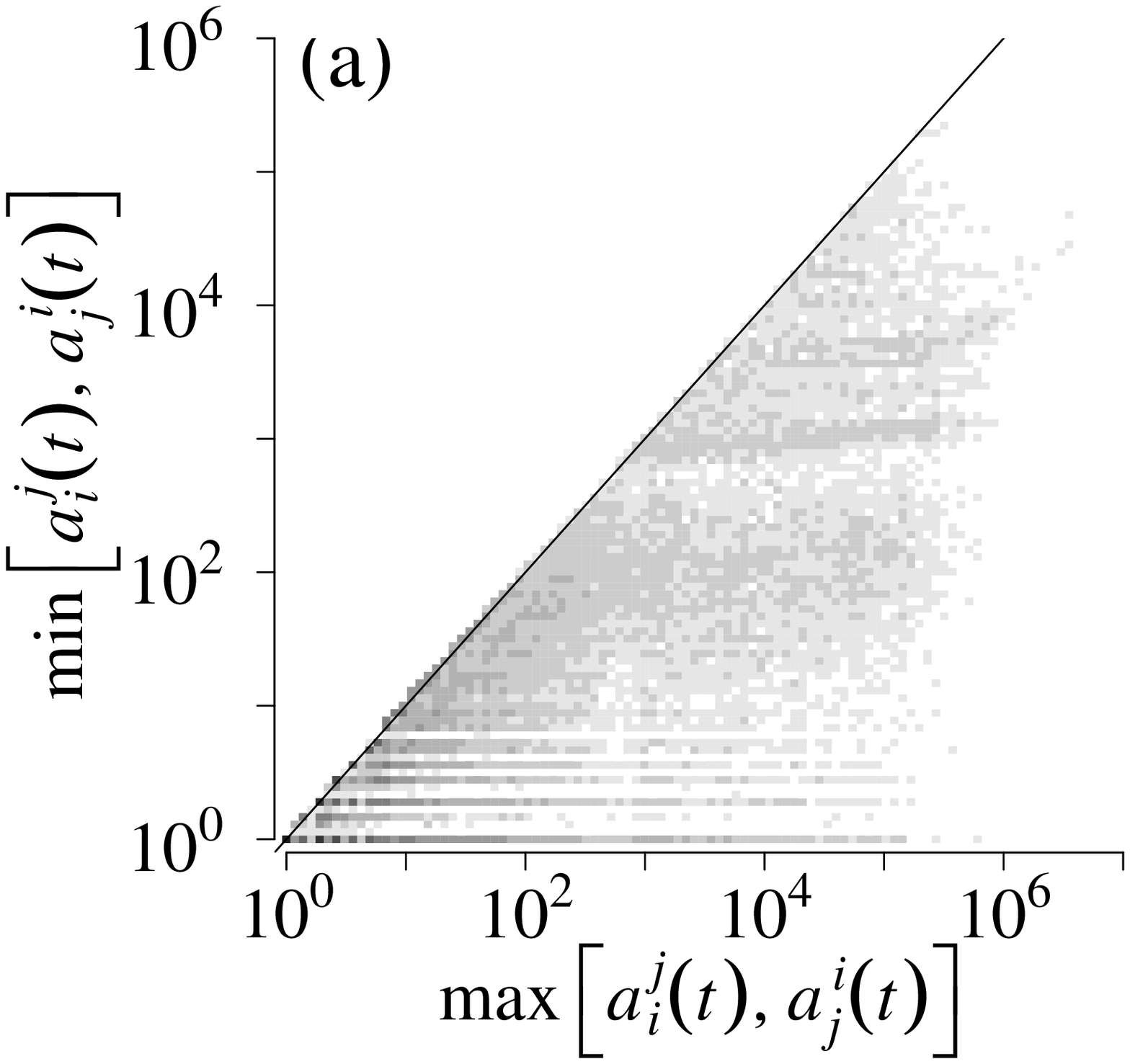}
\includegraphics[width=0.45\hsize, clip]{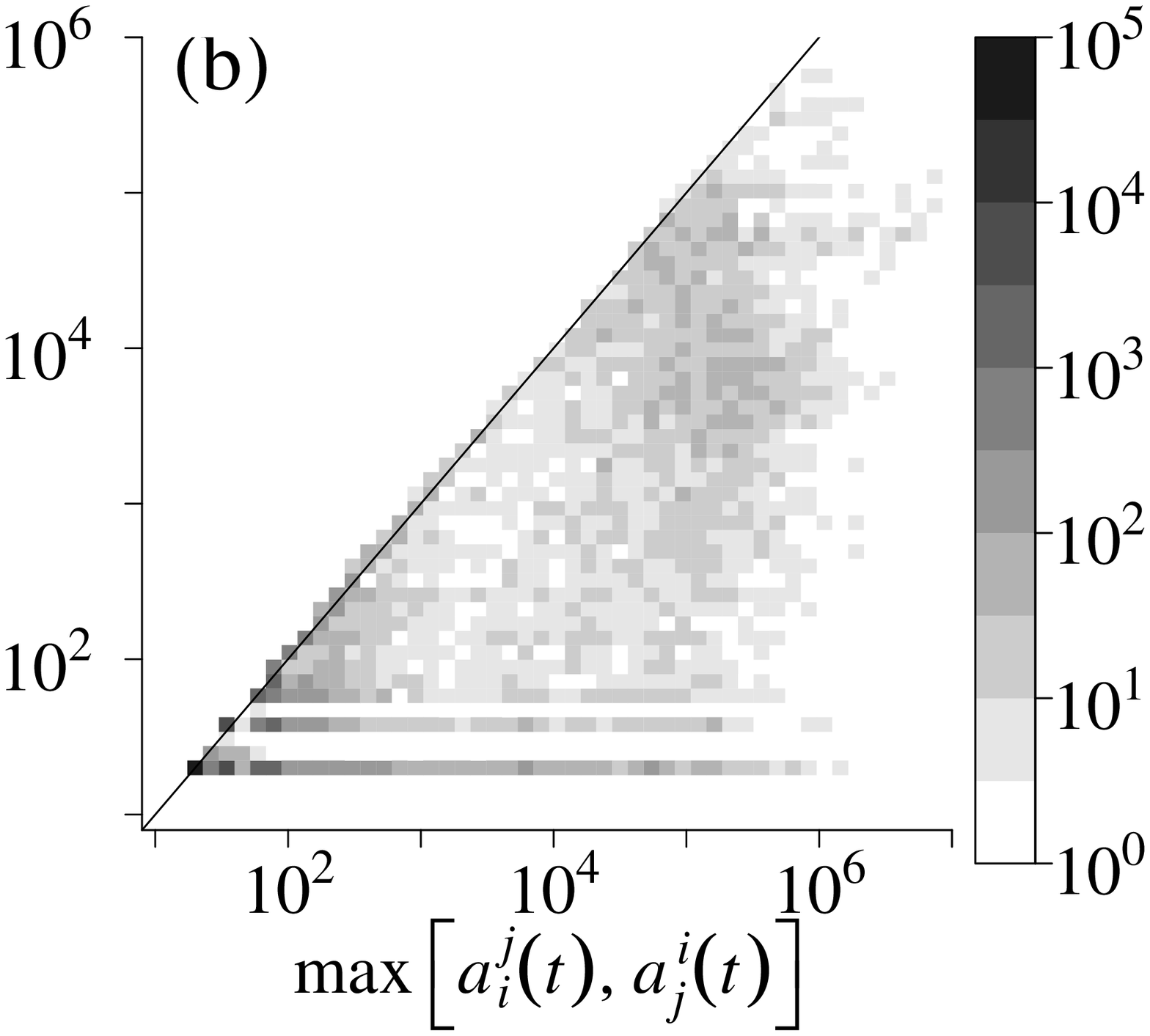}
\caption{Asymmetry in the advance of events for (a) Office1 and (b) Conference data sets.
The solid lines represent the diagonal on which $a_i^j(t)= a_j^i(t)$.}
\label{fig:asym_advance}
\end{figure}

\clearpage

\clearpage
\begin{figure}
\centering
\includegraphics[width=0.4\hsize,clip]{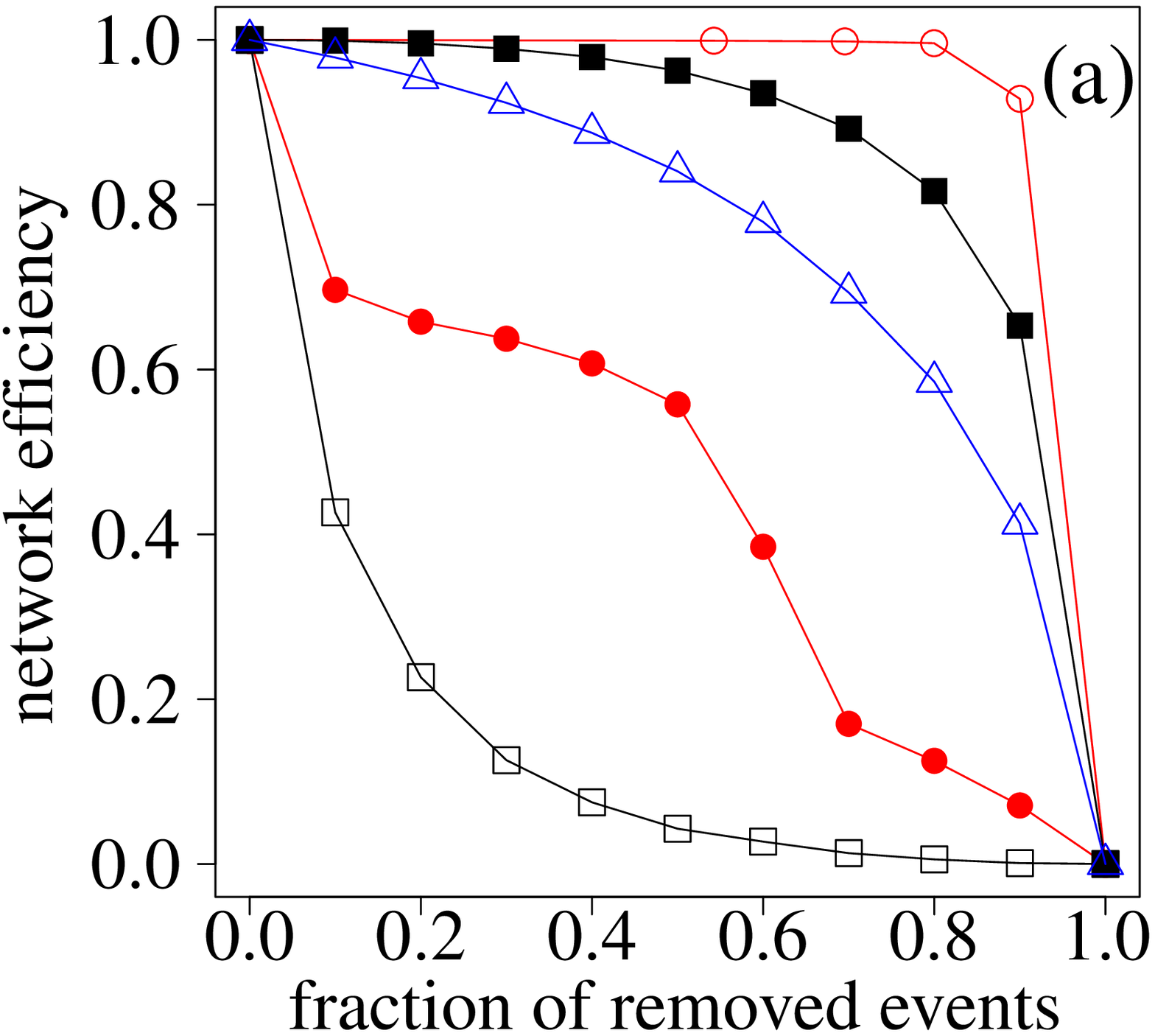}
\includegraphics[width=0.4\hsize,clip]{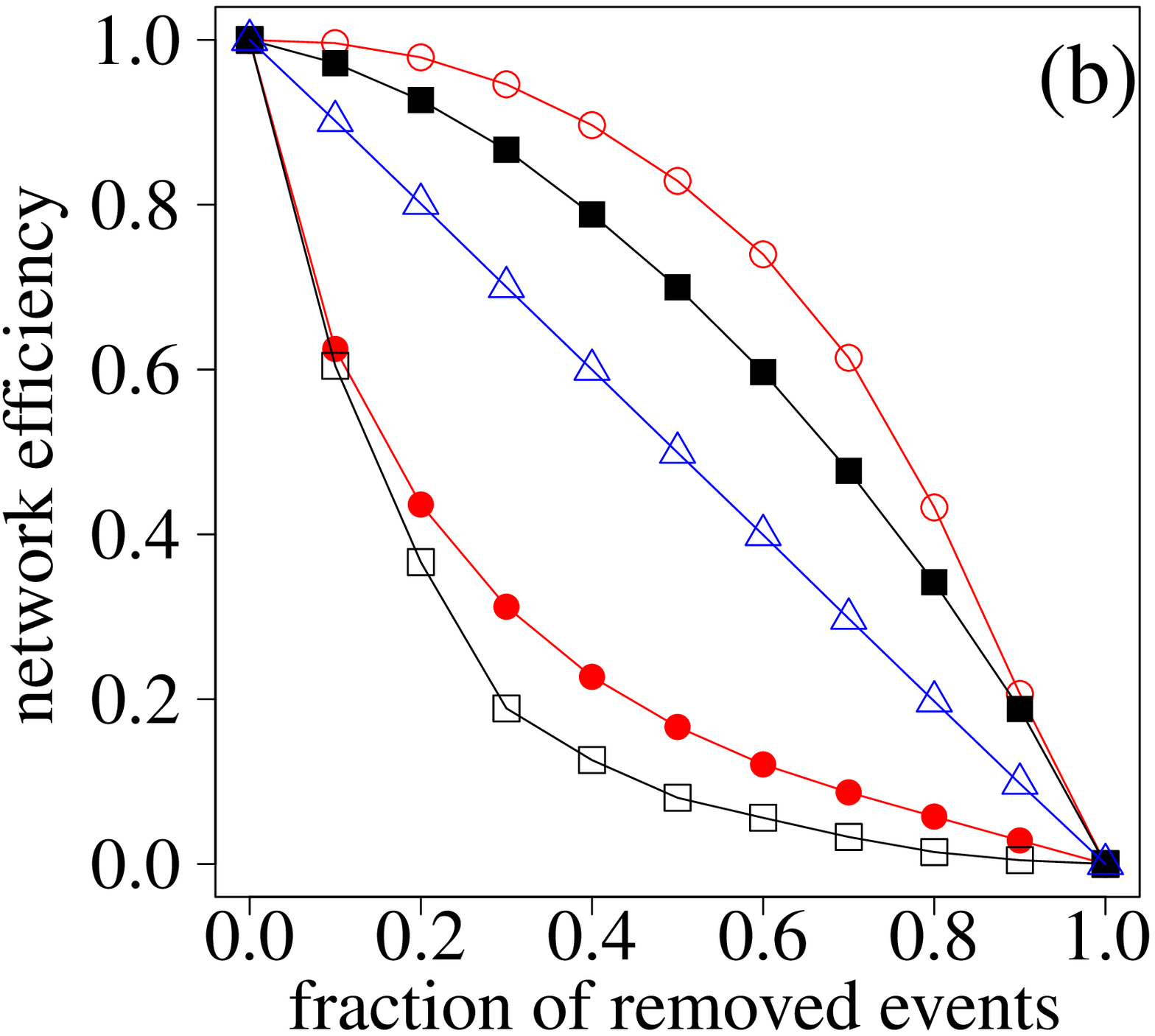}\\
\includegraphics[width=0.4\hsize,clip]{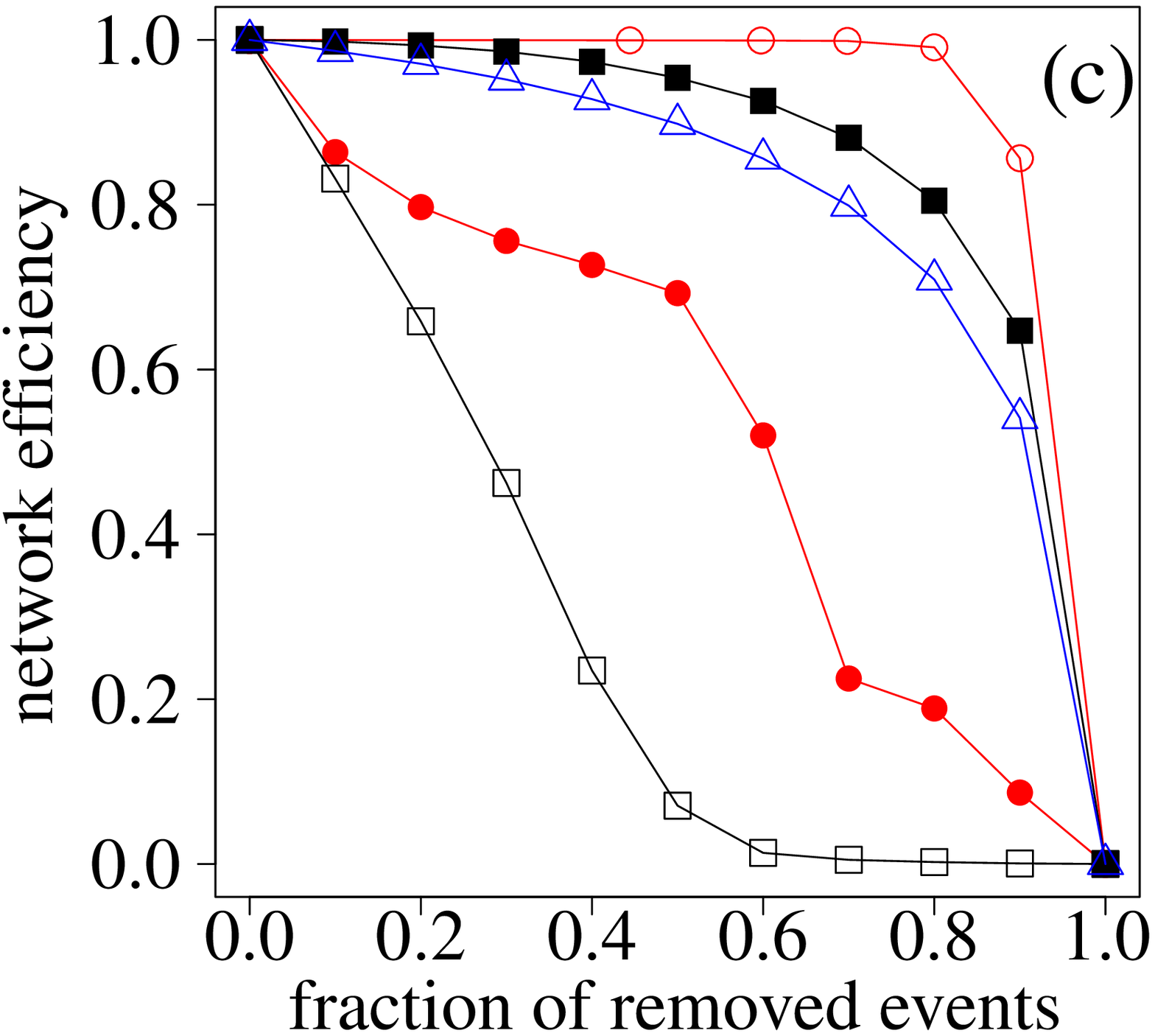}
\includegraphics[width=0.4\hsize,clip]{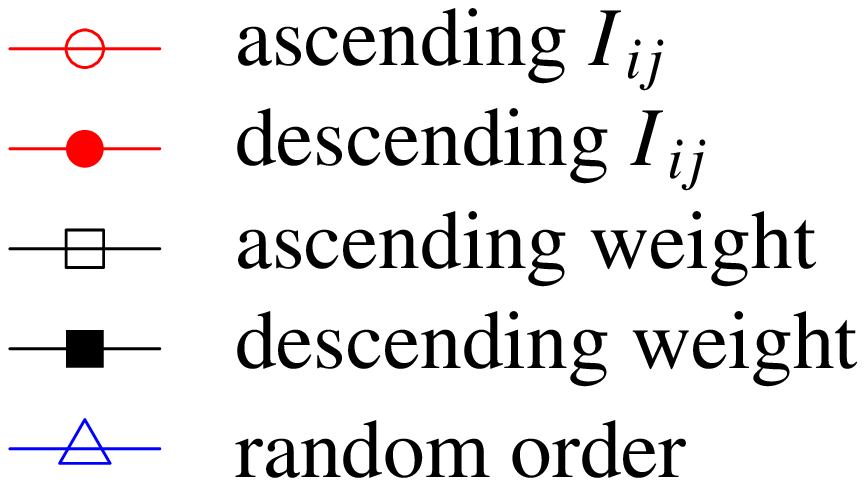}
\caption{Network efficiency for the randomized temporal networks generated from Office1 data set.
We generated the randomized temporal networks by (a) shuffling the IEIs, (b) Poissonizing the IEIs, and (c) randomly rewiring the links.
}
\label{fig:E_RI_RT_RG_event_removal}
\end{figure}

\clearpage

\clearpage
\setcounter{figure}{0}
\renewcommand{\thefigure}{A\arabic{figure}}
\begin{figure}
\centering
\includegraphics[width=0.4\hsize,clip]{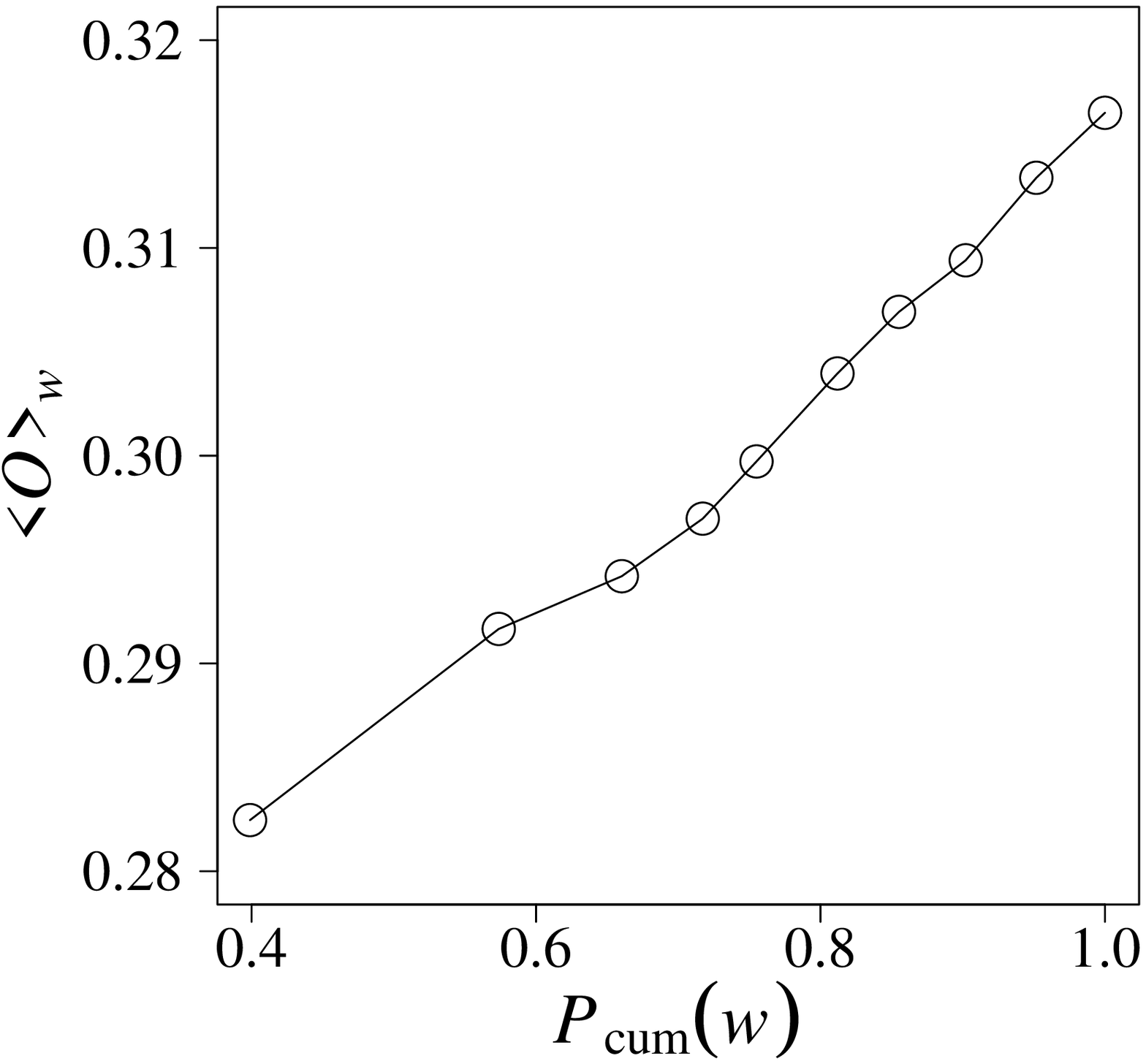}
\caption{Averaged neighborhood overlap $\langle O \rangle_w$ plotted against the fraction of links with weights smaller than or equal to $w$ for Conference data set.}
\label{fig:Oij_conference}
\end{figure}

\clearpage
\begin{table}
\centering
\caption{Statistics for the three data sets.}
\label{tab:data_info}
\begin{tabular}{|l|r|r|r|}
\hline
 & \makebox[5em][c]{Office1} & \makebox[5em][c]{Office2} & \makebox[5em][c]{Conference}\\\hline 
Number of individuals ($N$) & 163 & 211 & 113\\\hline
Total number of events & 118,456 & 274,308 & 20,818\\\hline
Observation period (day) & 73 & 120 & 3\\\hline
Time resolution & 1 min & 1 min & 20 sec\\\hline
\end{tabular}
\end{table}

\clearpage
\begin{table}
\centering
\caption{
Properties of temporal networks that are conserved and dismissed by the different randomizations.
$\surd$ and $-$ indicate conserved and dismissed, respectively.}
\label{tab:randomization}
\begin{tabular}{|l|c|c|c|}
\hline
 & \makebox[7em][c]{Shuffled IEIs} & \makebox[7em][c]{Poissonized IEIs} & \makebox[7em][c]{Rewiring}\\\hline
Burstiness & $\surd$ & $-$ & $\surd$\\\hline
Network structure & $\surd$ & $\surd$ & $-$\\\hline
Temporal correlation & $-$ & $-$ & $\surd$\\\hline
Distribution of link weight & $\surd$ & $\surd$ & $\surd$\\\hline
\end{tabular}
\end{table}

\clearpage
\setcounter{page}{1}
\begin{center}
{\bf {\Large Supplementary data}}\\
\vspace*{3mm}
{\large for}\\
\vspace*{3mm}
\textit{{\large Taro Takaguchi, Nobuo Sato, Kazuo Yano, and Naoki Masuda}}\\
\vspace*{3mm}
{\bf {\large Importance of individual events in temporal networks}}\\
\end{center}
\clearpage

\setcounter{figure}{0}
\renewcommand{\thefigure}{S\arabic{figure}}

\clearpage
\begin{figure}
\centering
\includegraphics[width=0.4\hsize]{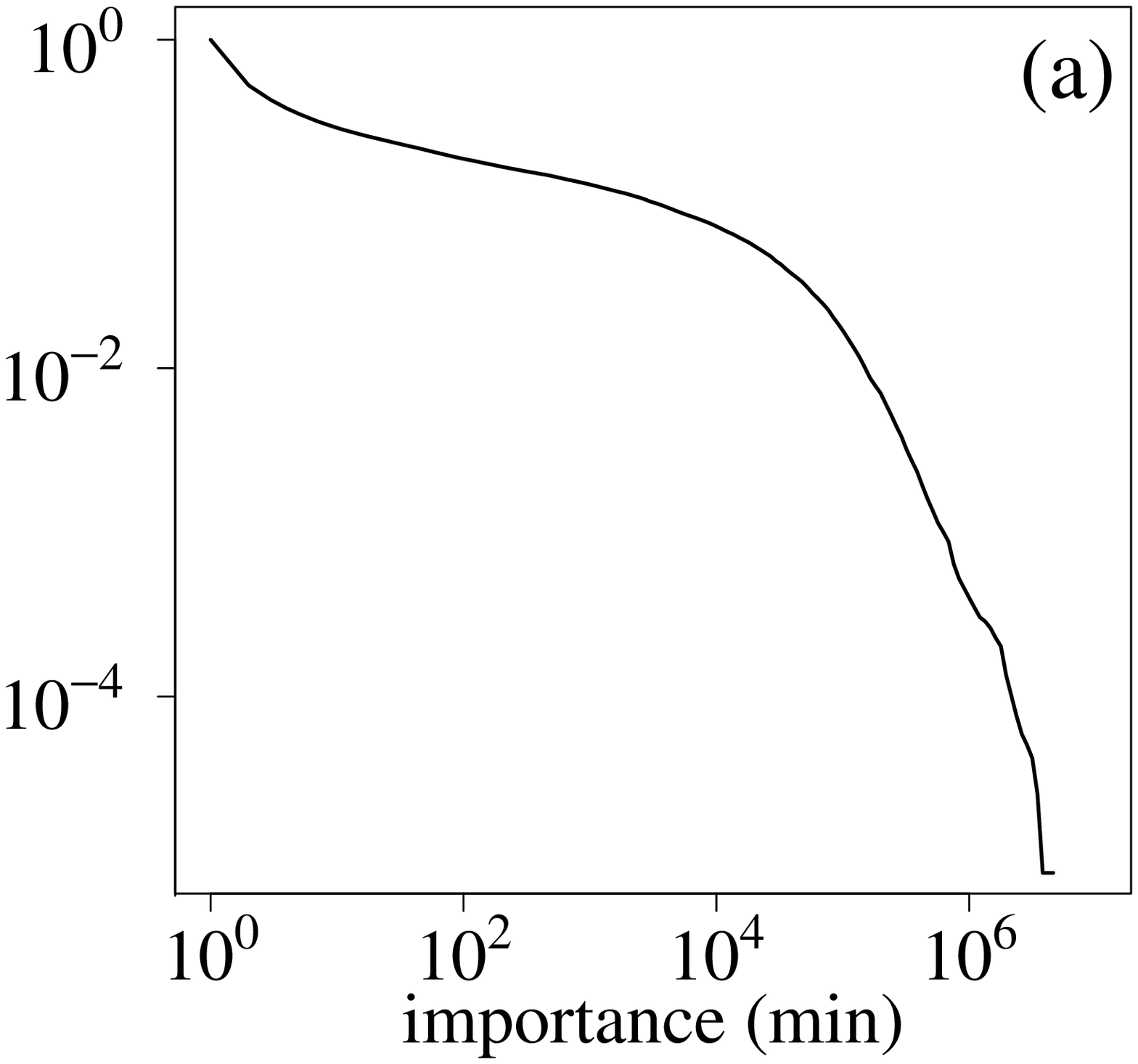}
\includegraphics[width=0.4\hsize]{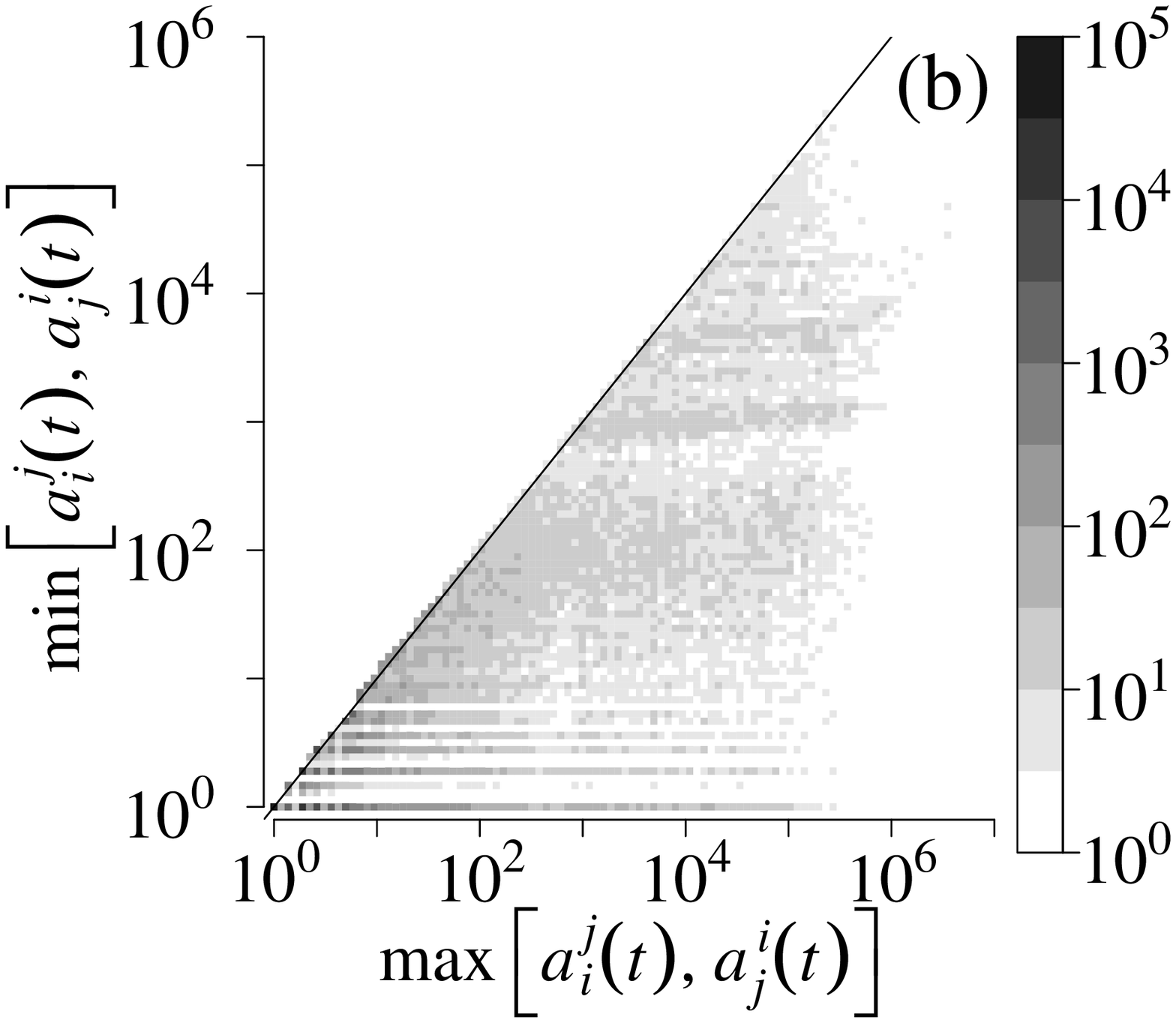}
\includegraphics[width=0.4\hsize]{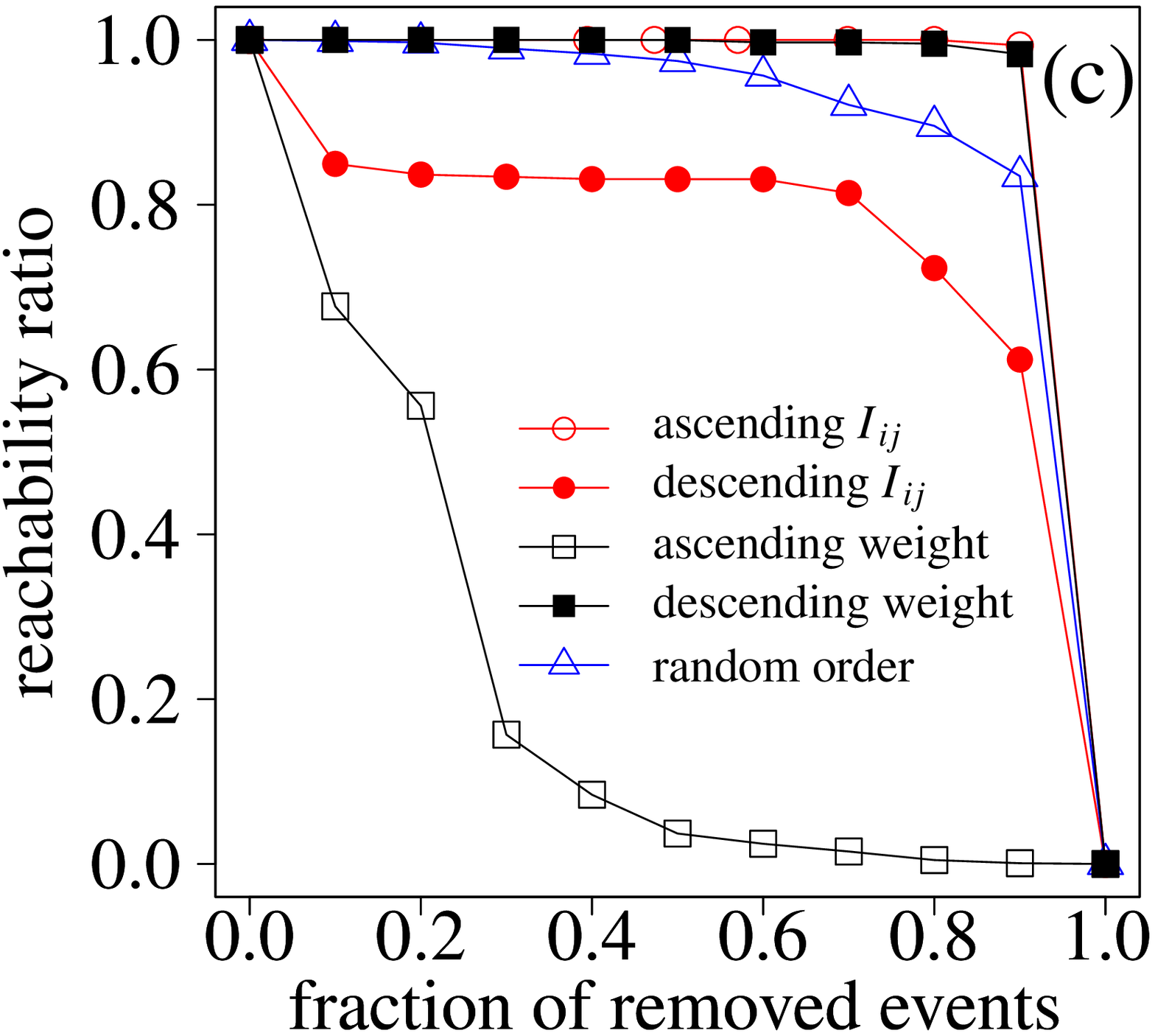}
\includegraphics[width=0.4\hsize]{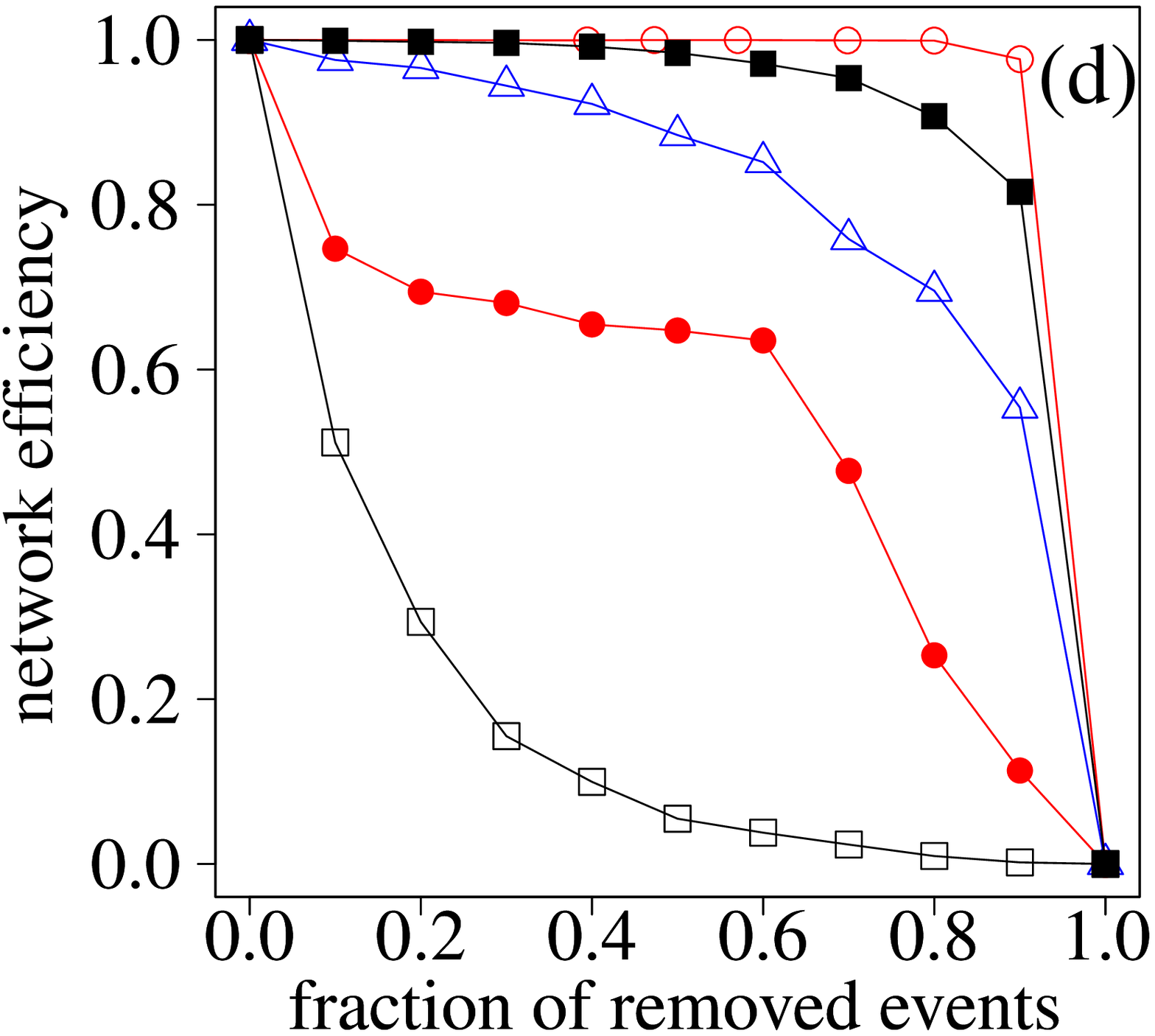}
\caption{
Results with $h=1$ for Office1 data set.
(a) Complementary cumulative distribution of the importance of event.
(b) Asymmetry in the advance of events.
(c) Reachability ratio. (d) Network efficiency.
}
\label{fig:h1}
\end{figure}

\clearpage
\begin{figure}
\centering
\includegraphics[width=0.4\hsize]{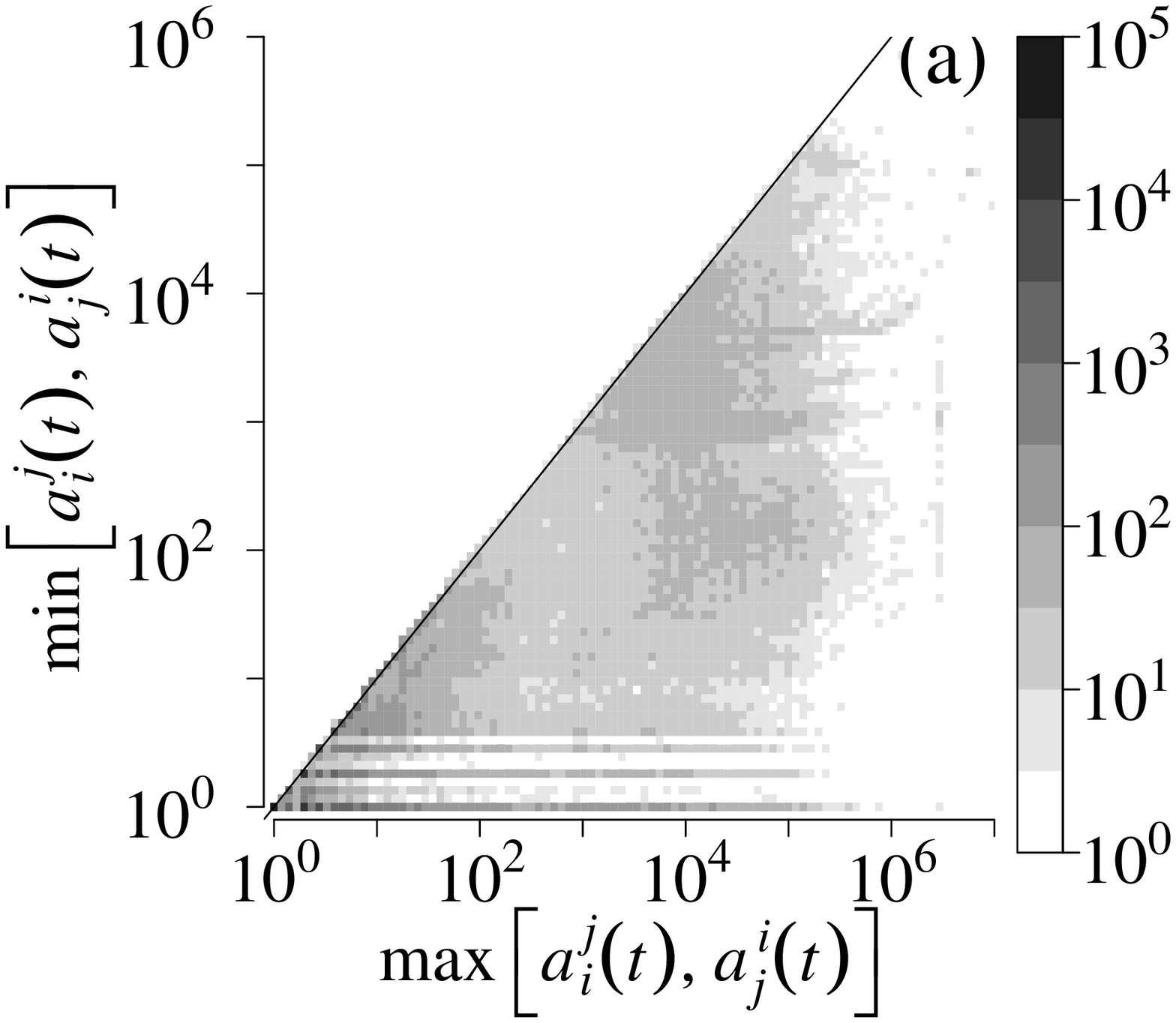}\\
\includegraphics[width=0.4\hsize]{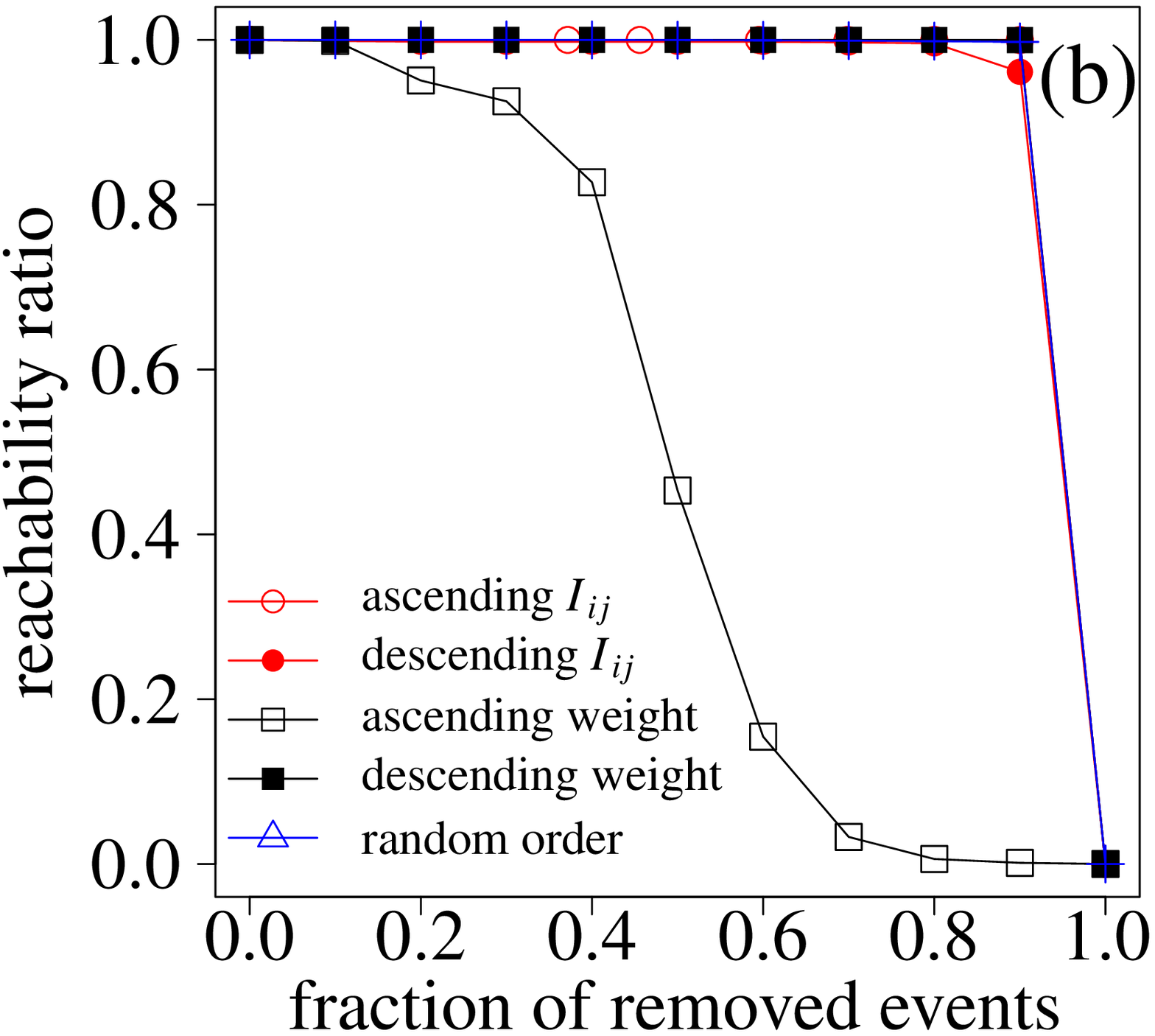}
\includegraphics[width=0.4\hsize]{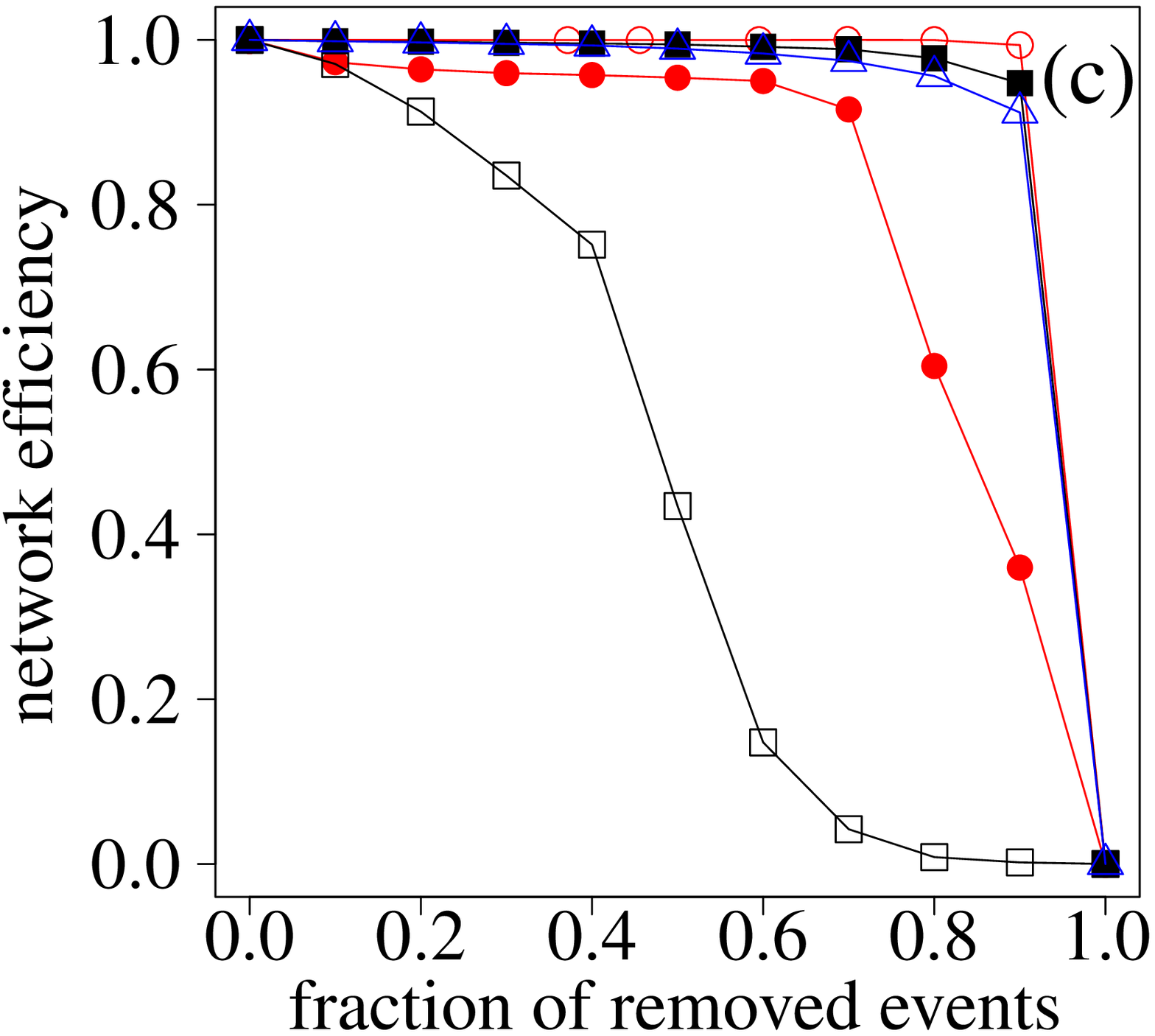}
\caption{
Results for Office2 data set.
(a) Asymmetry in the advance of events.
(b) Reachability ratio. (c) Network efficiency.
}
\label{fig:results_01}
\end{figure}

\clearpage
\begin{figure}
\centering
\includegraphics[width=0.4\hsize, clip]{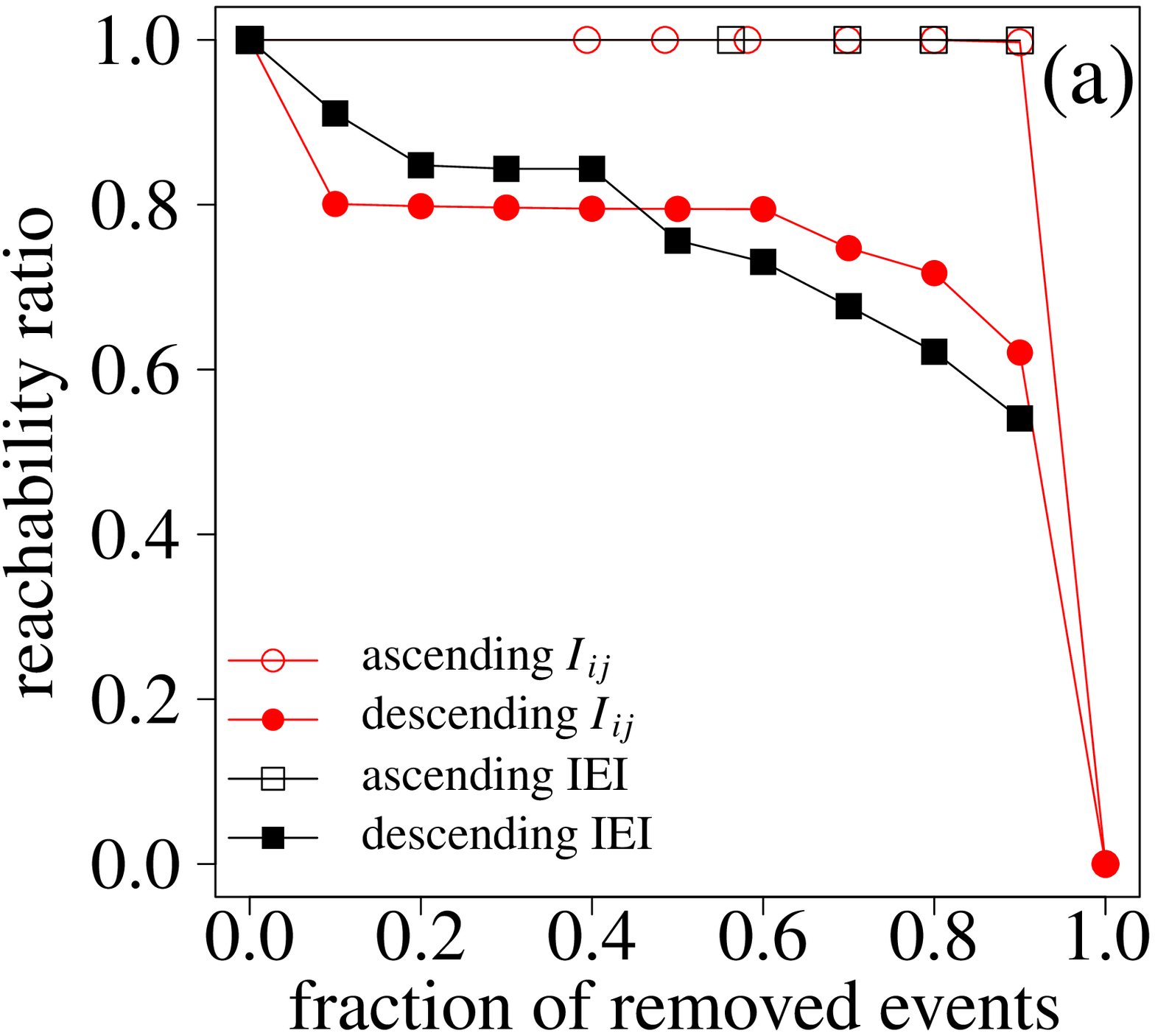}
\includegraphics[width=0.4\hsize, clip]{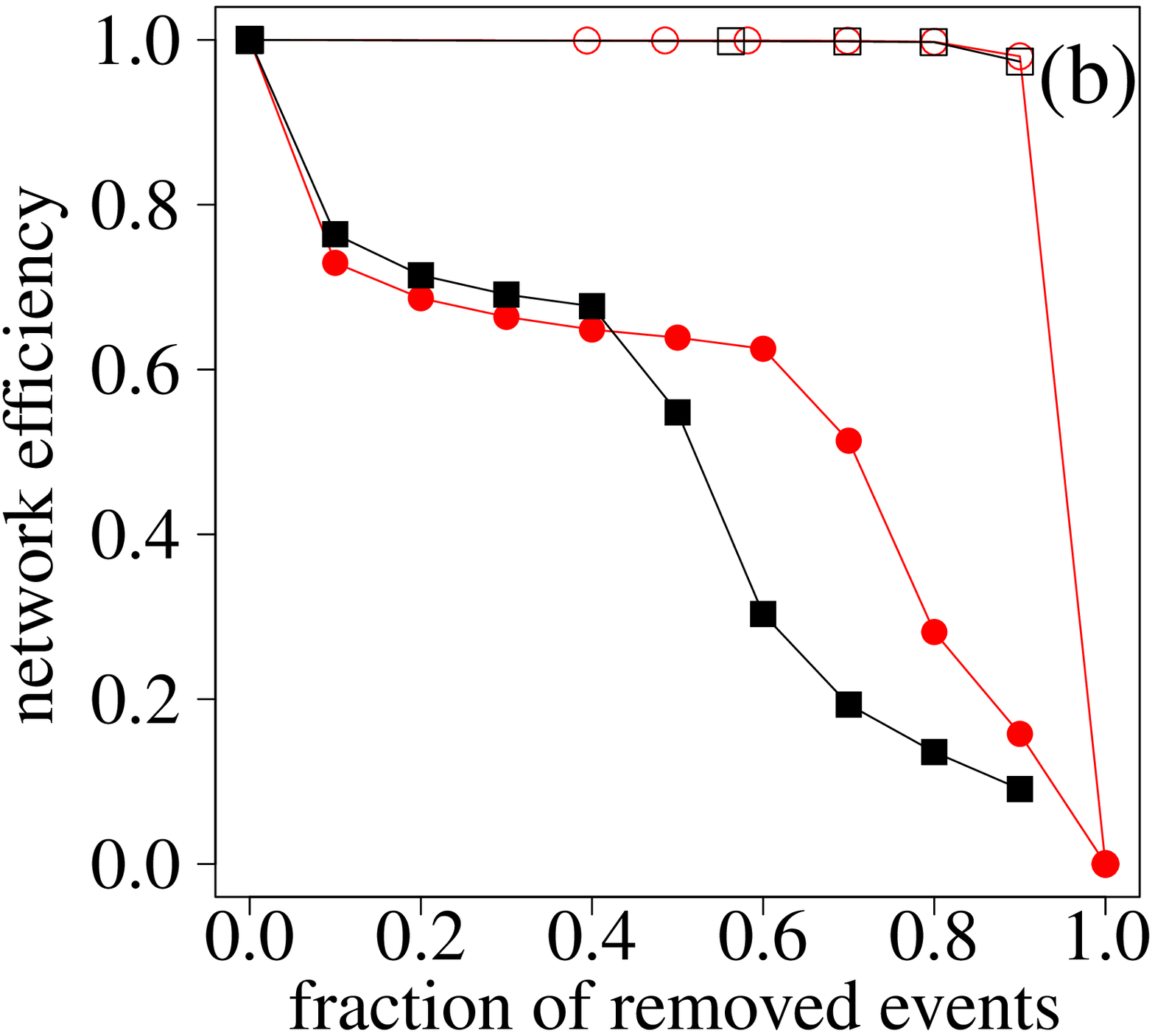}
\caption{Comparison of the event removal tests
on the basis of the importance of event and the length of the latest IEI before the event. We used Office1 data set.
(a) Reachability ratio. (b) Network efficiency.
}
\label{fig:event_removal_IEI_00}
\end{figure}

\clearpage
\begin{figure}
\centering
\includegraphics[width=0.4\hsize,clip]{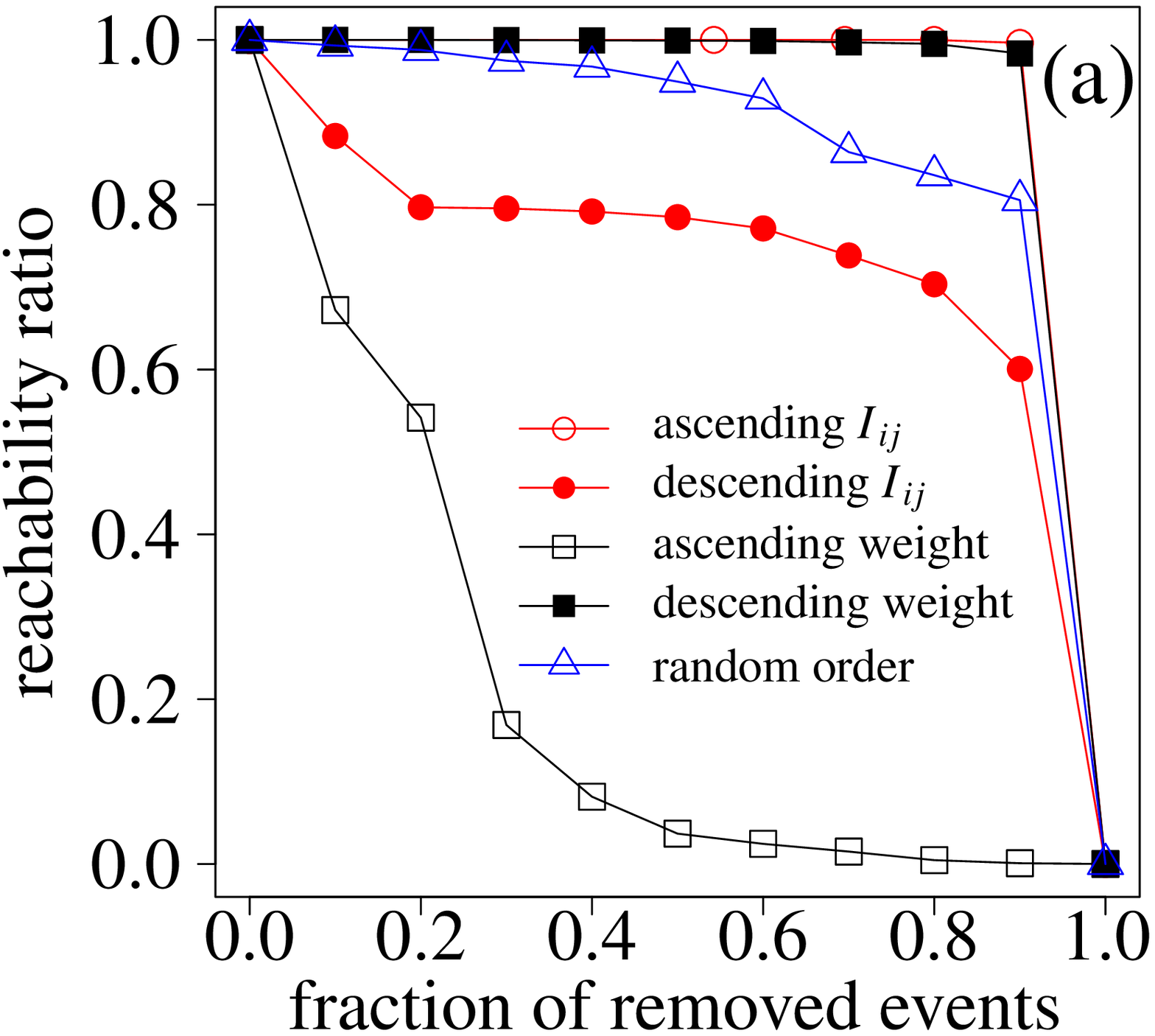}
\includegraphics[width=0.4\hsize,clip]{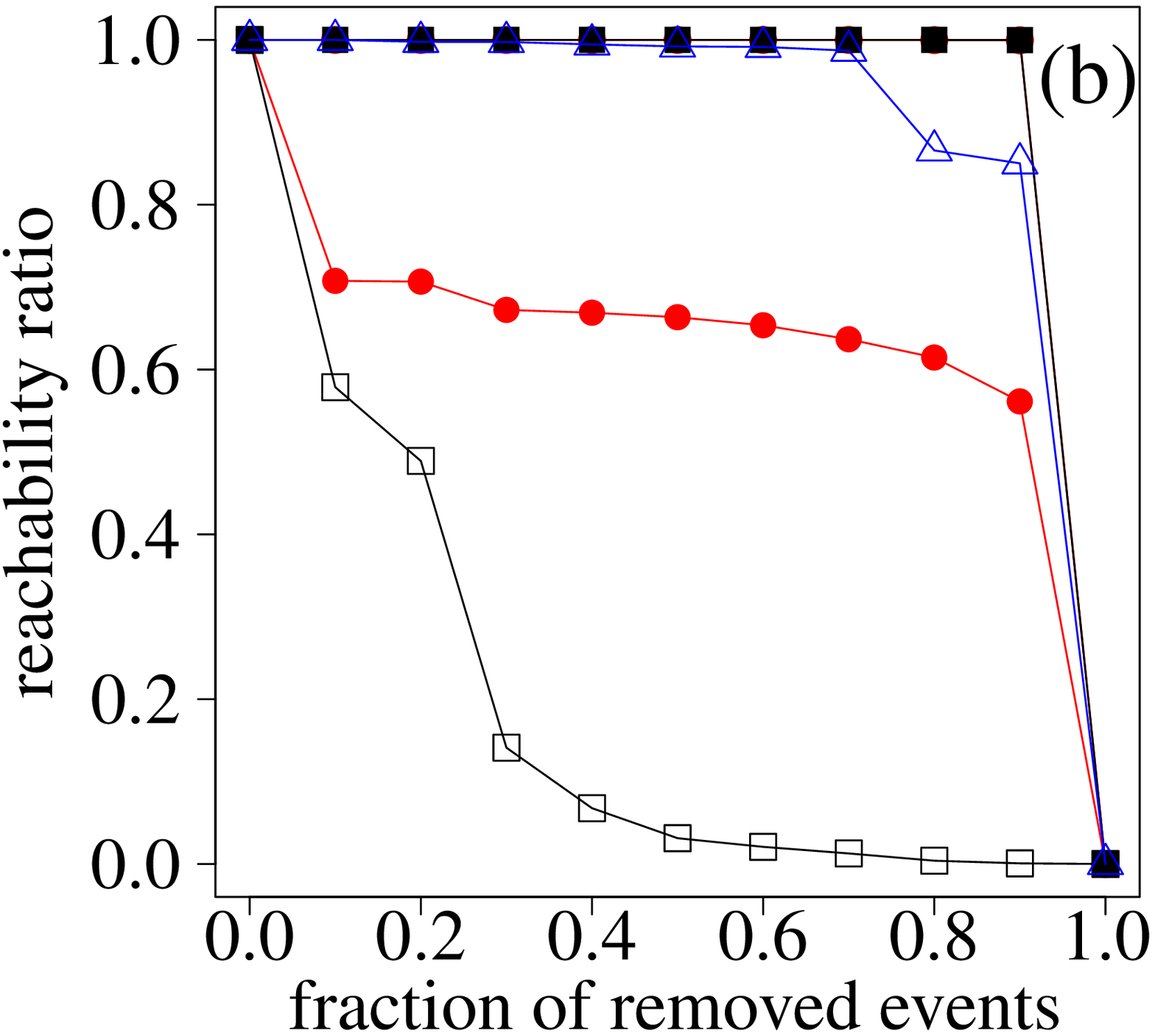}\\
\includegraphics[width=0.4\hsize,clip]{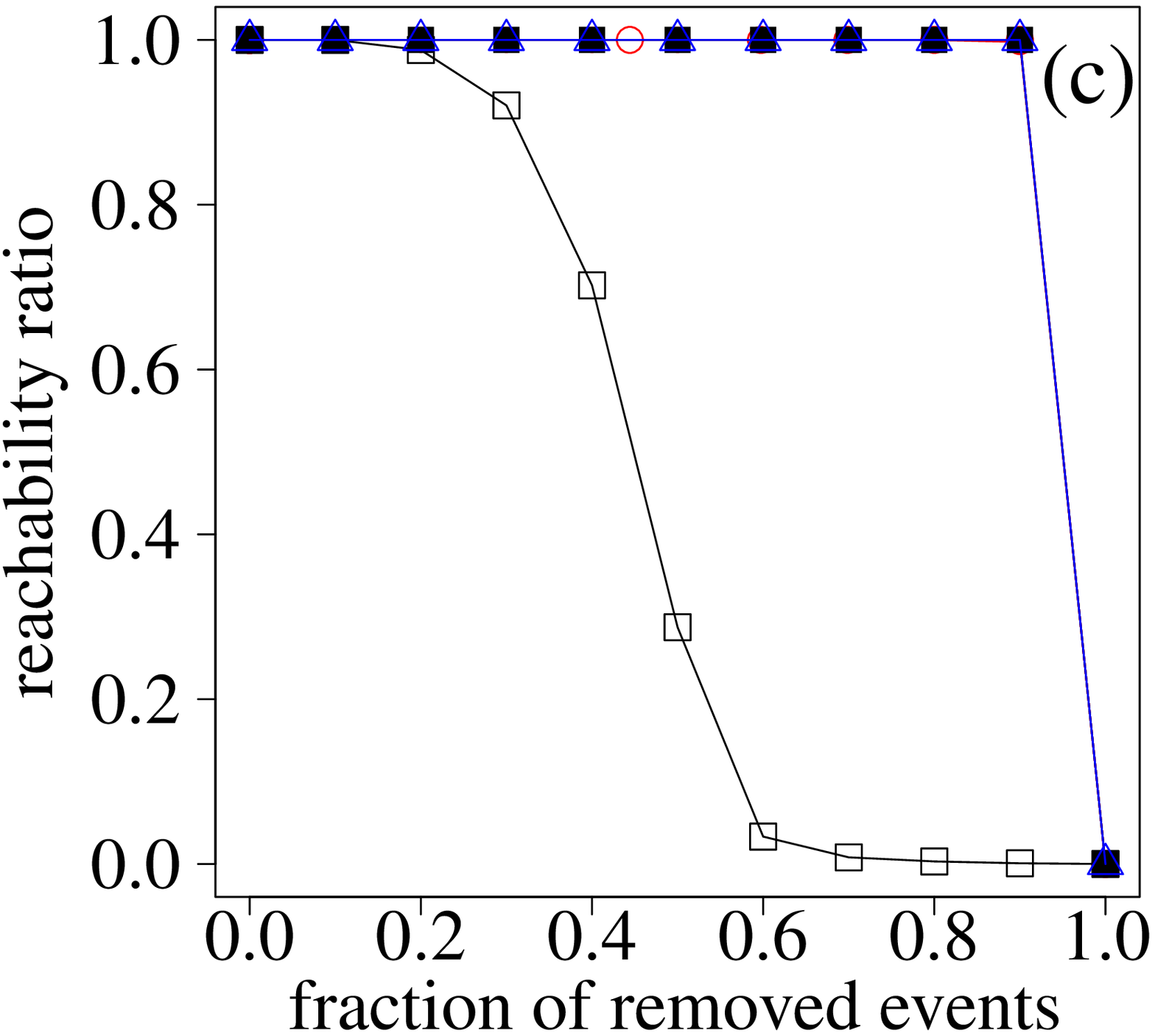}
\caption{Reachability ratio for the randomized temporal networks generated from Office1 data set.
We generated the randomized temporal networks by (a) shuffling the IEIs, (b) Poissonizing the IEIs, and (c) randomly rewiring the links.
}
\label{fig:f_RI_RT_RG_event_removal}
\end{figure}

\clearpage
\begin{figure}
\centering
\includegraphics[width=0.35\hsize,clip]{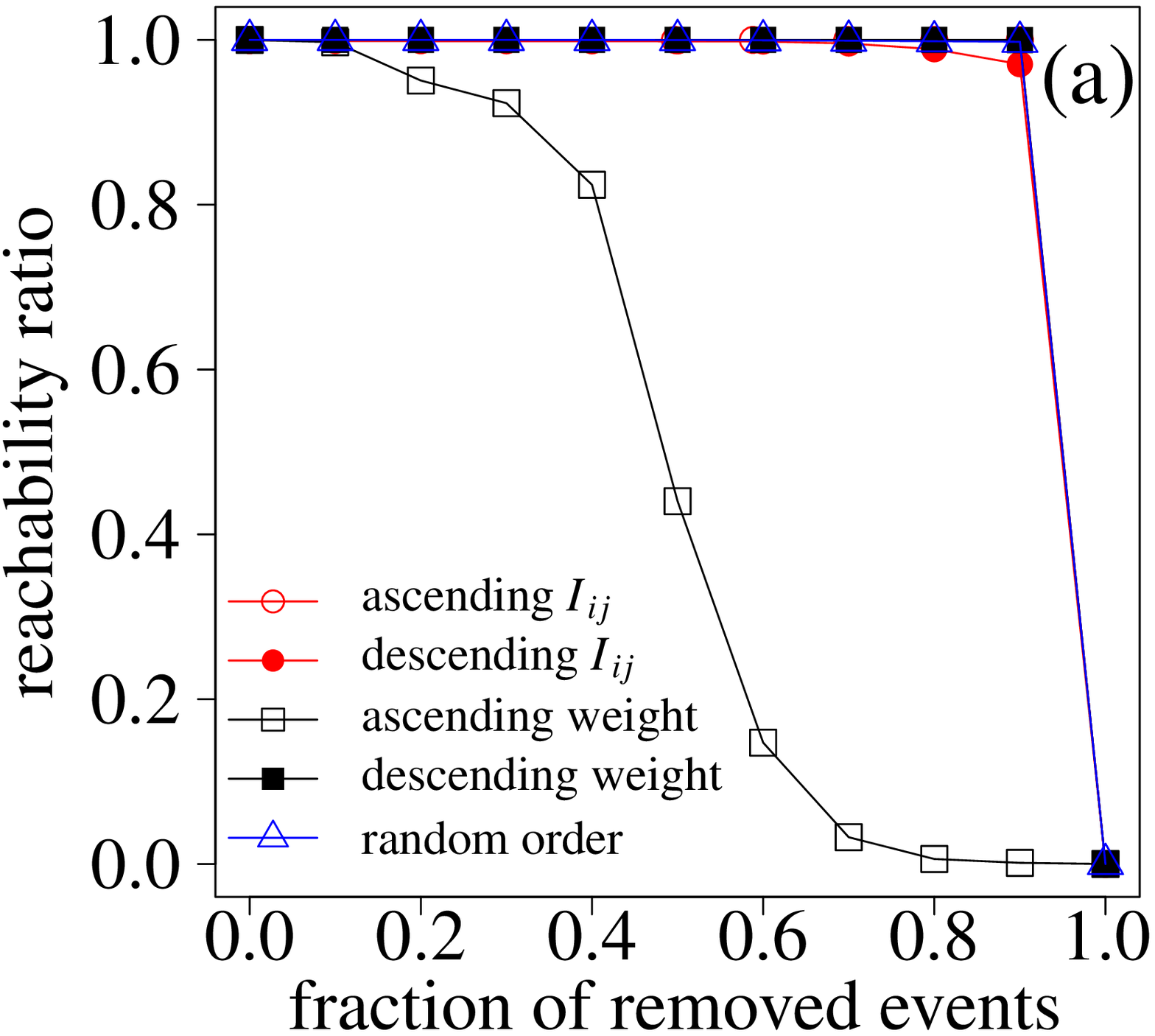}
\includegraphics[width=0.35\hsize,clip]{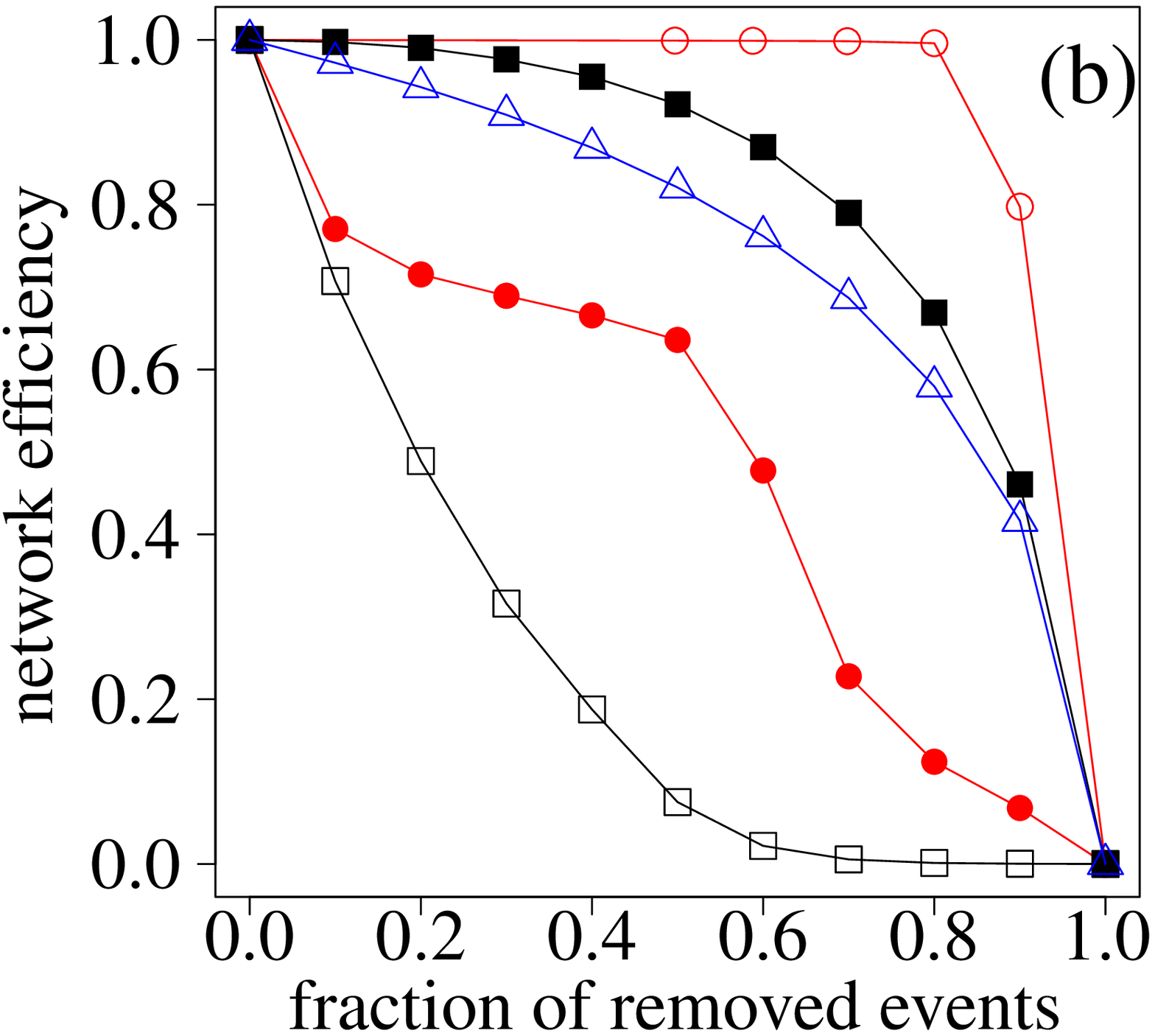}\\
\includegraphics[width=0.35\hsize,clip]{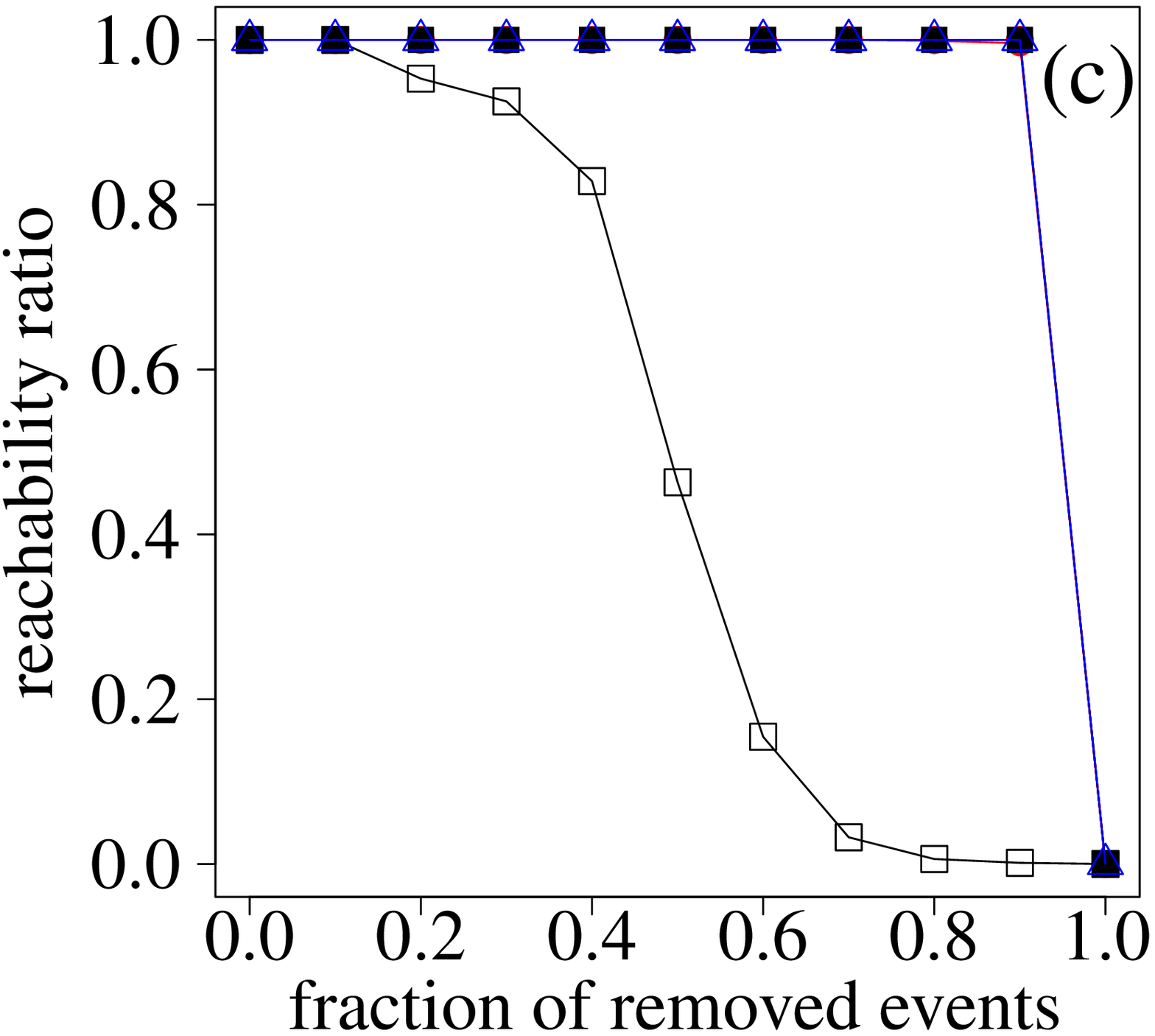}
\includegraphics[width=0.35\hsize,clip]{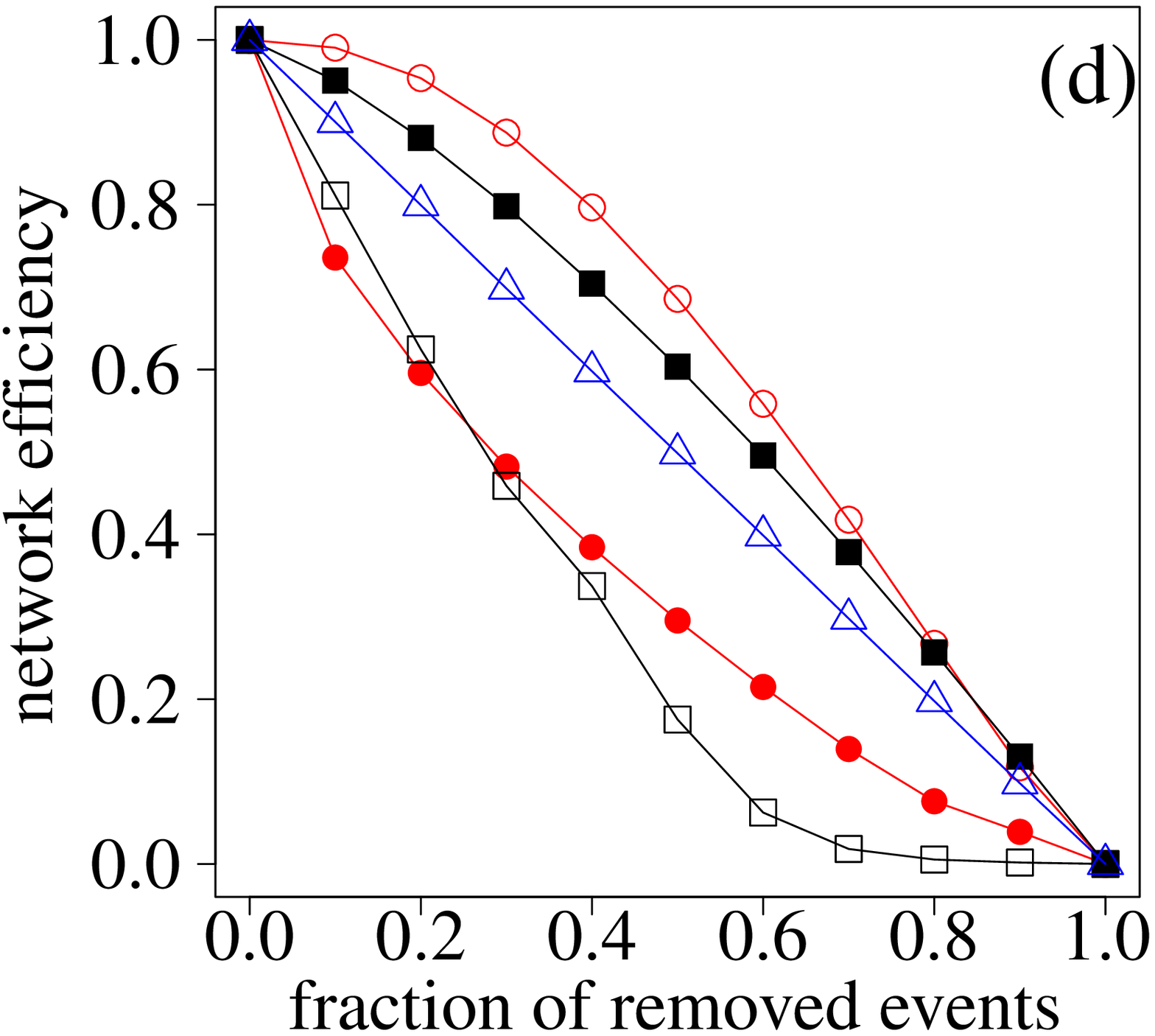}\\
\includegraphics[width=0.35\hsize,clip]{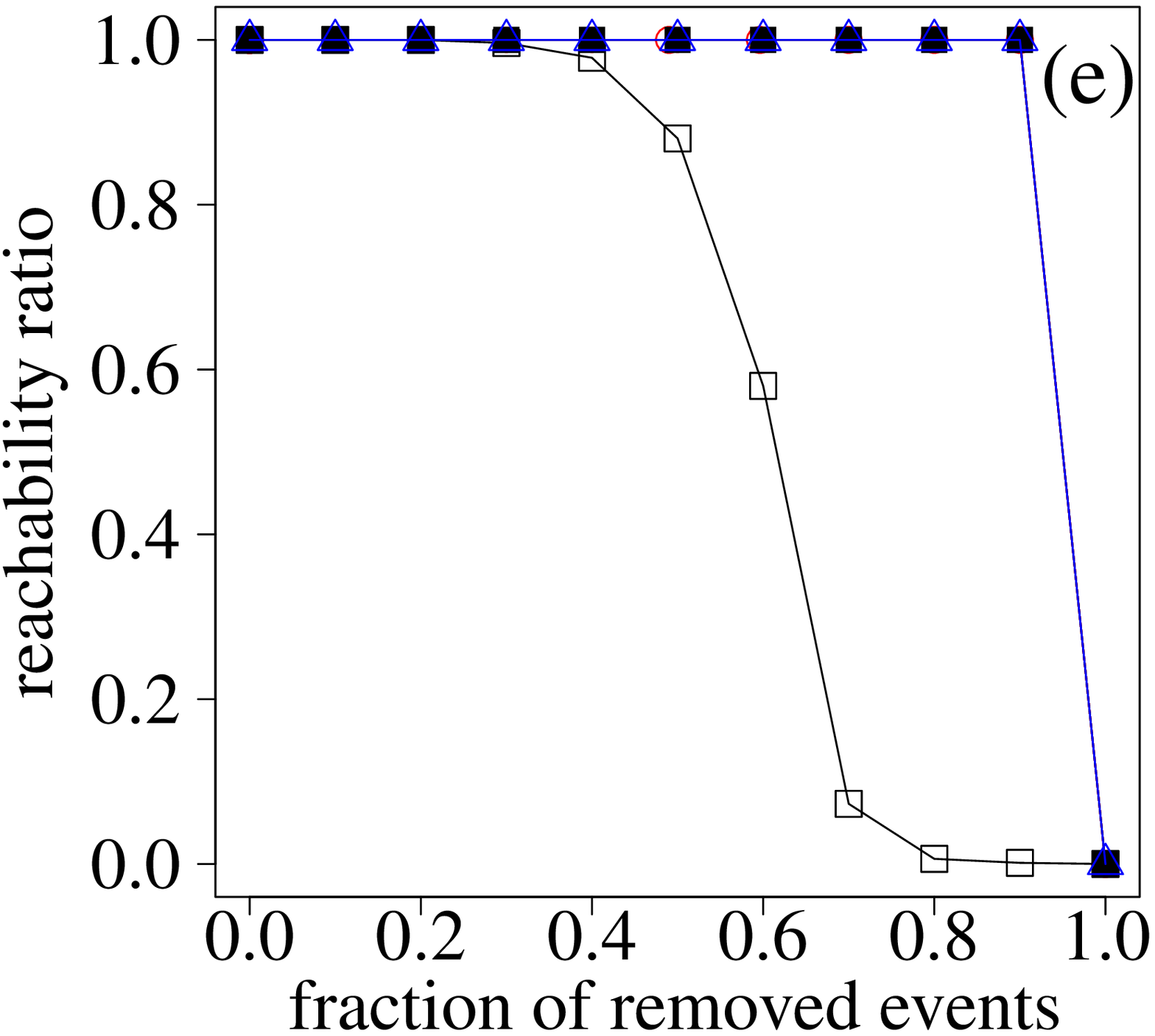}
\includegraphics[width=0.35\hsize,clip]{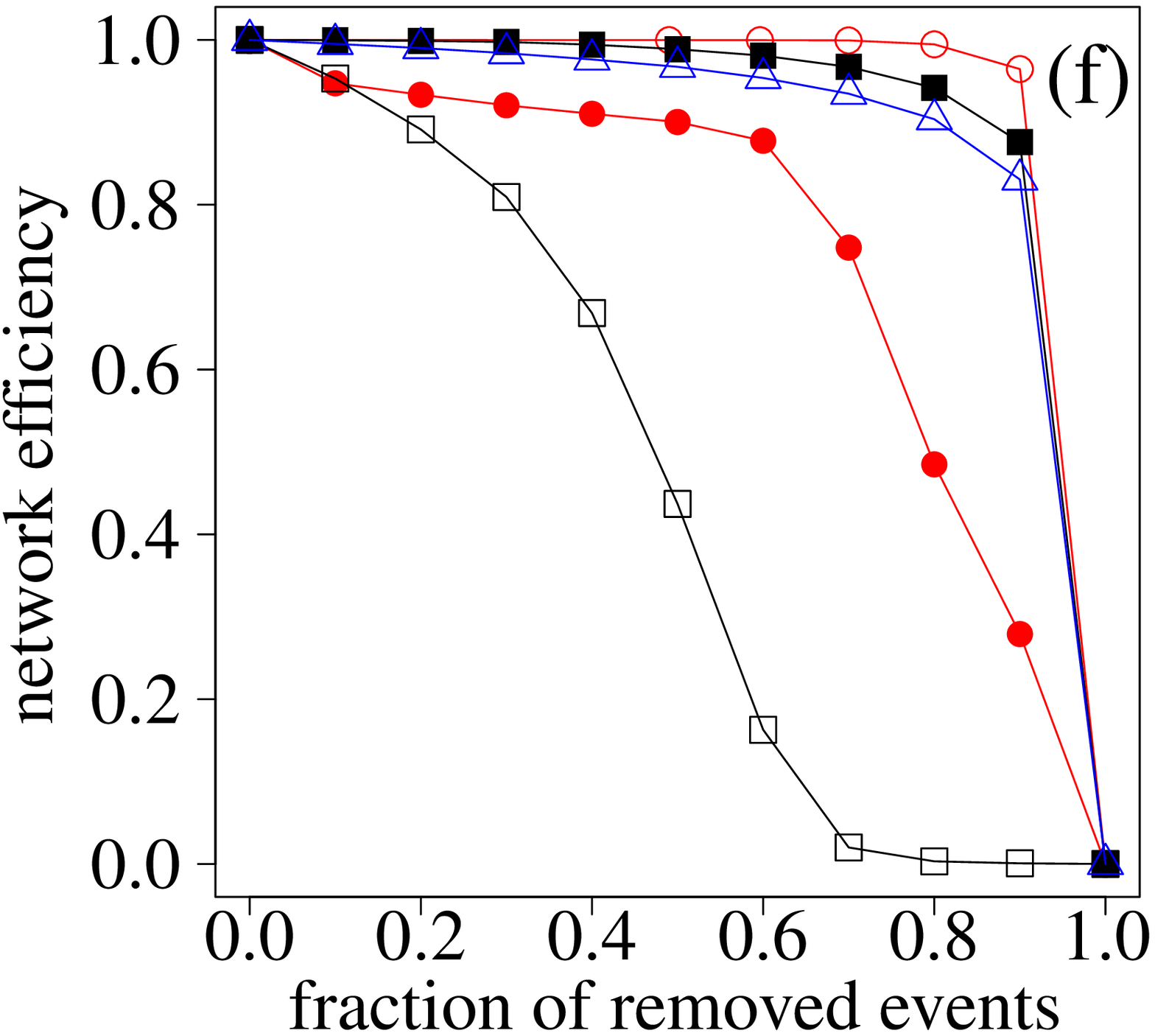}
\caption{
Results of the event removal tests for the randomized temporal networks generated from Office2 data set.
(a), (c), (e) Reachability ratio and (b), (d), (f) network efficiency. 
We generated the randomized temporal networks by (a), (b) shuffling the IEIs, (c), (d) Poissonizing the IEIs, and (e), (f) randomly rewiring the links.
}
\label{fig:RI_RT_RG_event_removal_01}
\end{figure}

\clearpage
\begin{figure}
\centering
\includegraphics[width=0.35\hsize,clip]{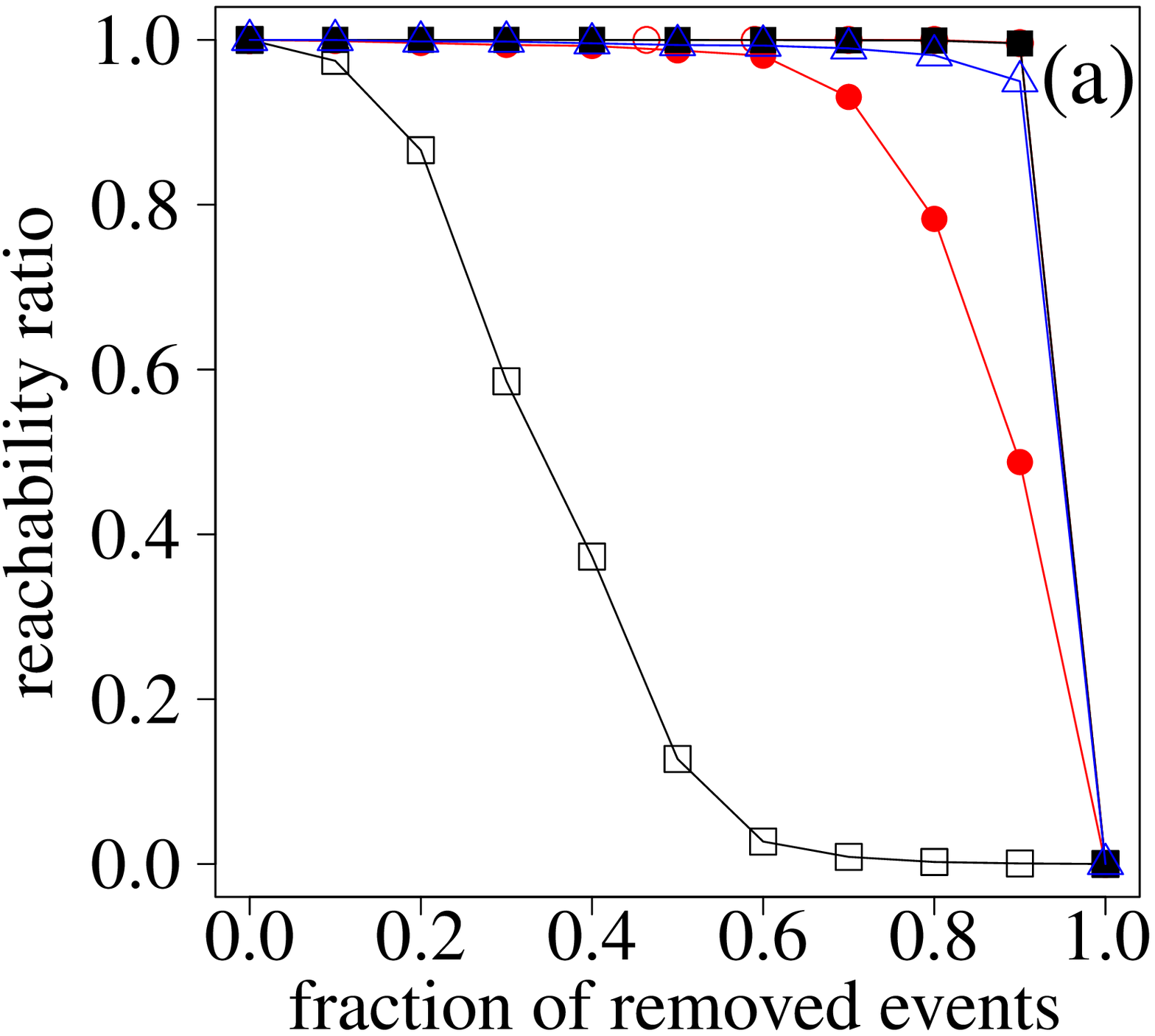}
\includegraphics[width=0.35\hsize,clip]{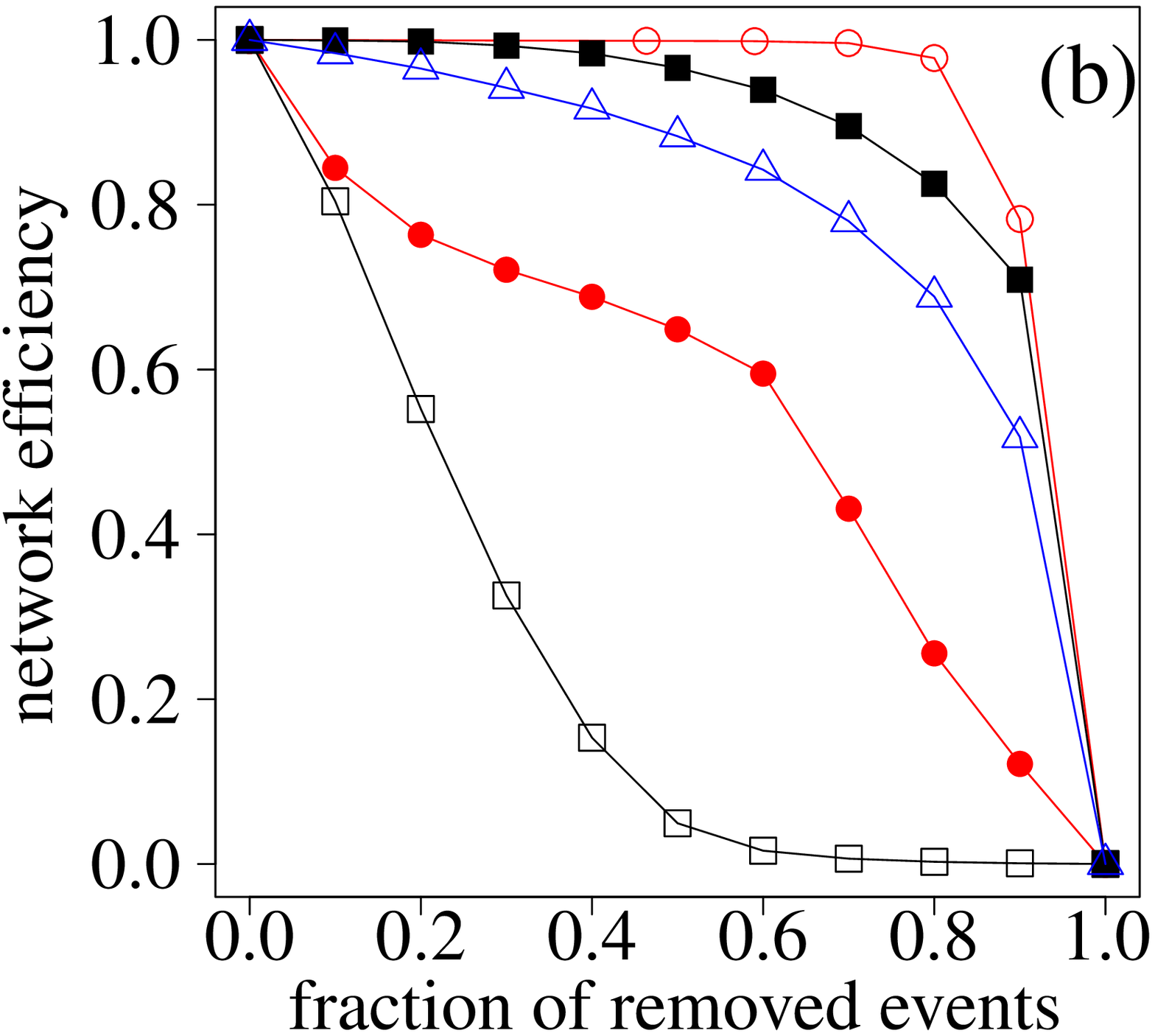}\\
\includegraphics[width=0.35\hsize,clip]{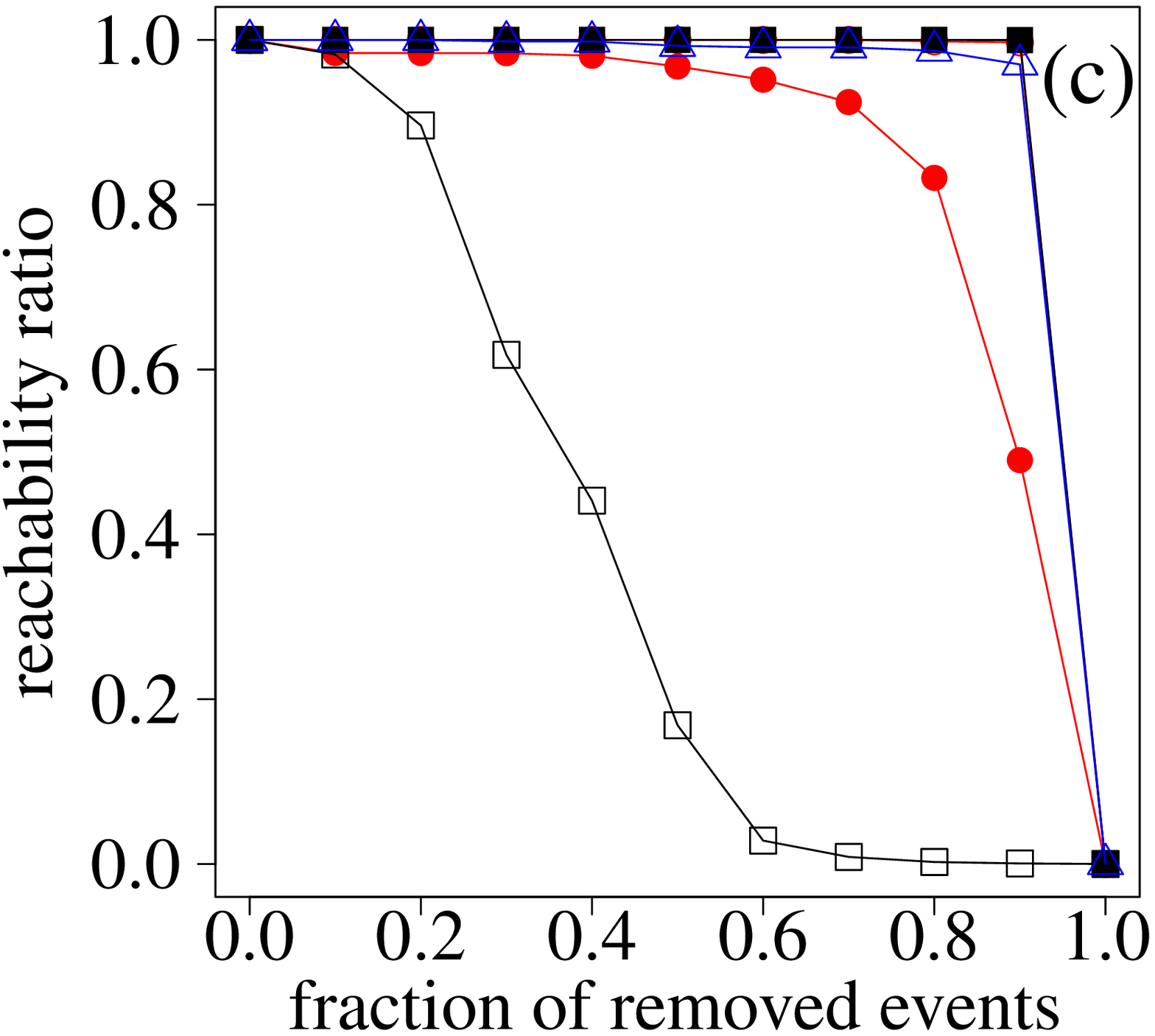}
\includegraphics[width=0.35\hsize,clip]{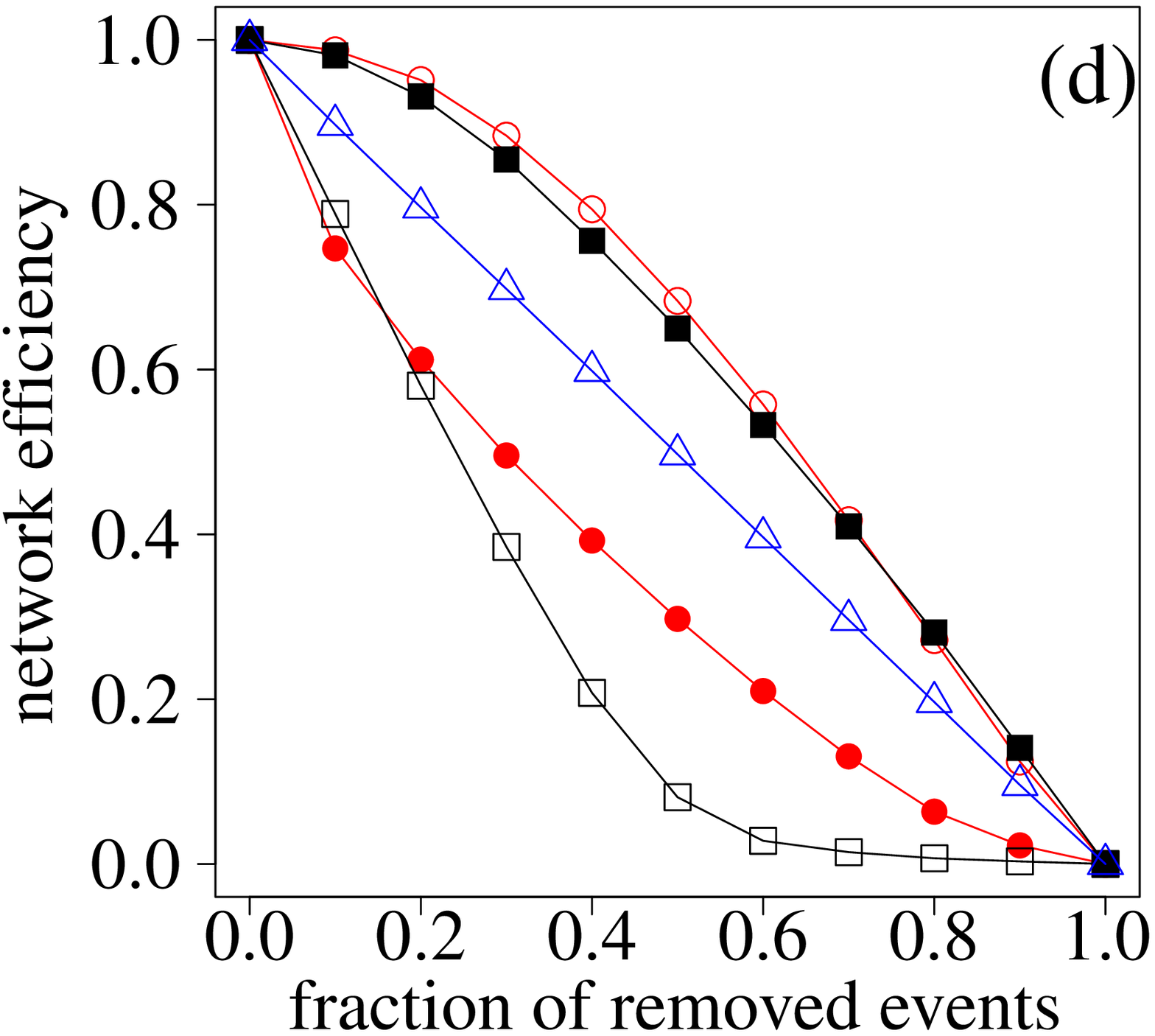}\\
\includegraphics[width=0.35\hsize,clip]{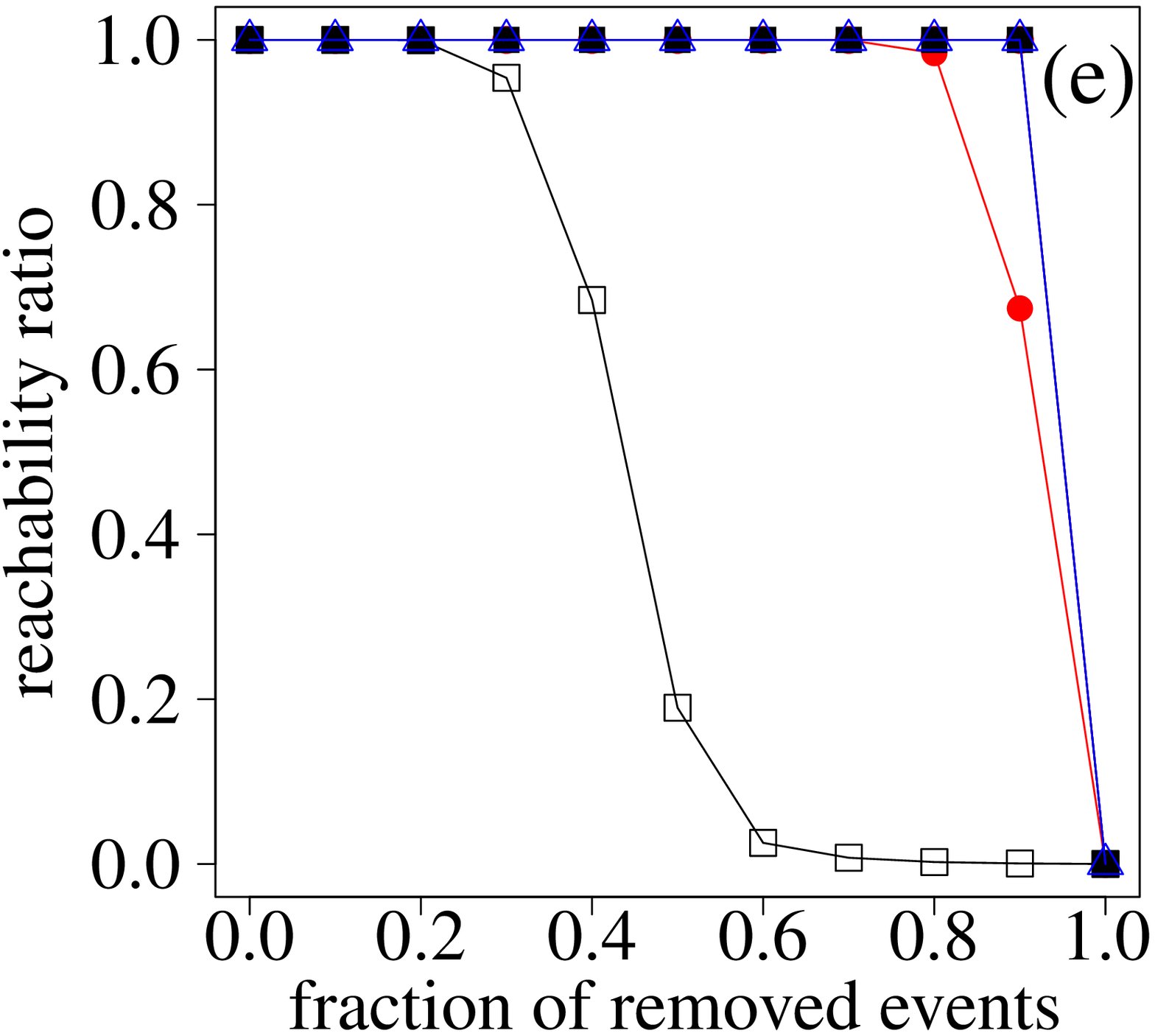}
\includegraphics[width=0.35\hsize,clip]{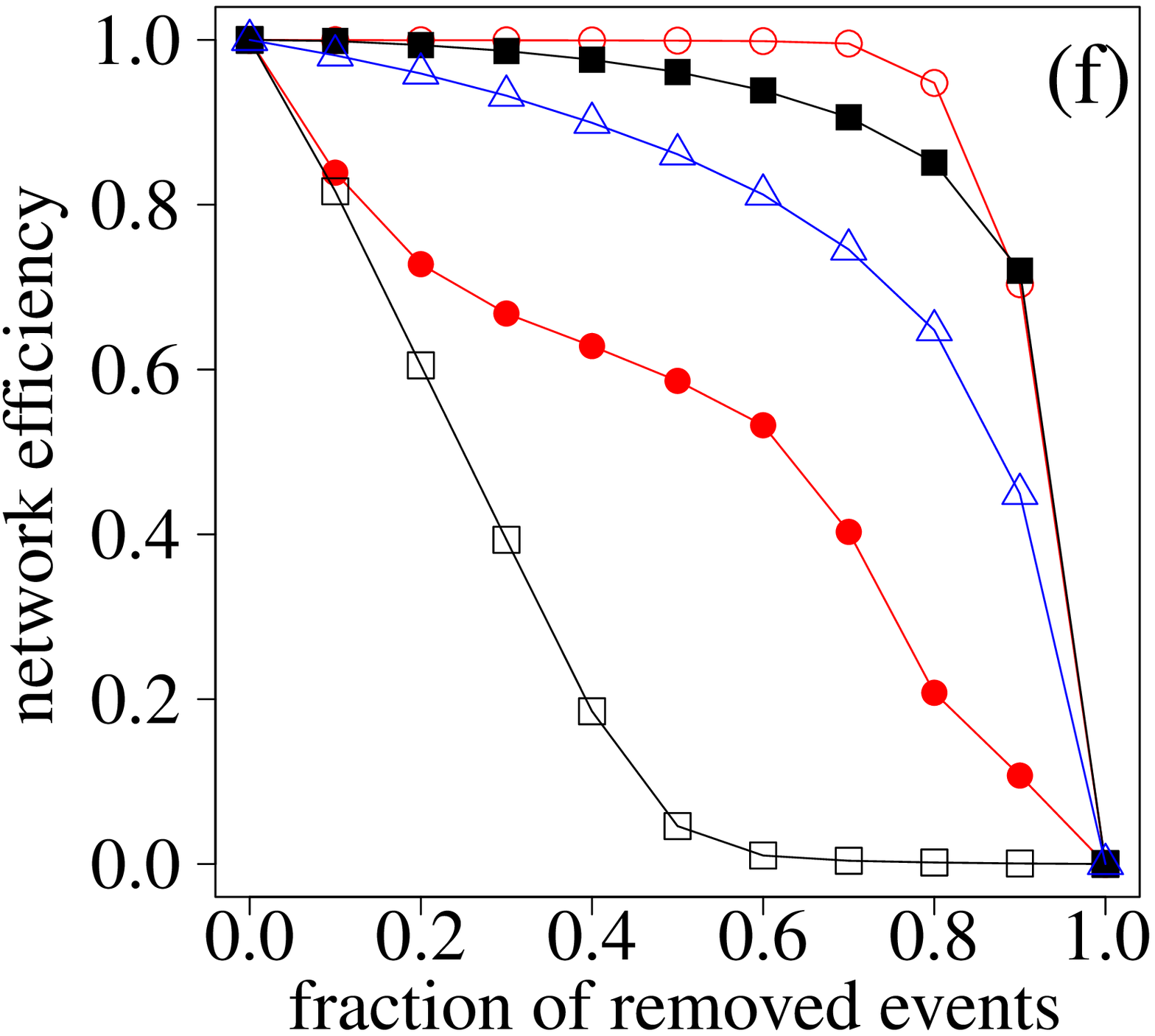}
\caption{
Results of the event removal tests for the randomized temporal networks generated from Conference data set.
(a), (c), (e) Reachability ratio and (b), (d), (f) network efficiency. 
We generated the randomized temporal networks by (a), (b) shuffling the IEIs, (c), (d) Poissonizing the IEIs, and (e), (f) randomly rewiring the links.
See the legend of supplementary figure~\ref{fig:RI_RT_RG_event_removal_01}(a) for the description of the symbols. 
}
\label{fig:RI_RT_RG_event_removal_HT09}
\end{figure}

\end{document}